\documentclass[aps,pra,twocolumn,showpacs,superscriptaddress,10pt,letterpaper,floatfix]{revtex4-1}
\usepackage{graphicx}% Include figure files
\usepackage{dcolumn}% Align table columns on decimal point
\usepackage{bm}% bold math
\usepackage{color}% add textcolor
\usepackage{amsmath}% for subequations environment
\usepackage{mdframed}
\usepackage{enumitem}
\usepackage[caption=false]{subfig}
\usepackage{mathtools}
\usepackage{tabularx}
\usepackage{multirow}
\usepackage{appendix}
\usepackage{longtable}
\usepackage{makecell}
\usepackage{hyperref}
\usepackage{multirow}

\definecolor{red}{rgb}{1,0,0}
\definecolor{orange}{rgb}{1,0.5,0}
\definecolor{green}{rgb}{0.13,0.55,0.13}
\definecolor{purple}{rgb}{0.5,0,1}

\begin{document}

\title{Rubric-based holistic review represents a change from traditional graduate admissions approaches in physics}

  \author{Nicholas T. Young}
   \email[Current email: ]{ntyoung@umich.edu}
  \affiliation {Department of Physics and Astronomy, Michigan State University, East Lansing, Michigan 48824}
  \affiliation {Department of Computational Mathematics, Science, and Engineering, Michigan State University, East Lansing, Michigan 48824}
  
  \author{N. Verboncoeur}
   \affiliation {Department of Physics and Astronomy, Michigan State University, East Lansing, Michigan 48824}
   
   \author{Dao Chi Lam}
    \affiliation {Department of Statistics, Michigan State University, East Lansing, Michigan 48824}
    
  \author{Marcos D. Caballero}
  \email[Corresponding Author: ]{caball14@msu.edu}
  \affiliation {Department of Physics and Astronomy, Michigan State University, East Lansing, Michigan 48824}
  \affiliation {Department of Computational Mathematics, Science, and Engineering, Michigan State University, East Lansing, Michigan 48824}
  \affiliation {Center for Computing in Science Education \& Department of Physics, University of Oslo, N-0316 Oslo, Norway}
  \affiliation {CREATE for STEM Institute, Michigan State University, East Lansing, Michigan 48824}

\date{\today}

\begin{abstract}
Rubric-based admissions are claimed to help make the graduate admissions process more equitable, possibly helping to address the historical and ongoing inequities in the U.S. physics graduate school admissions process that have often excluded applicants from minoritized races, ethnicities, genders, and backgrounds. Yet, no studies have examined whether rubric-based admissions methods represent a fundamental change of the admissions process or simply represent a new tool that achieves the same outcome. To address that, we developed supervised machine learning models of graduate admissions data collected from our department over a seven-year period. During the first four years, our department used a traditional admissions process and switched to a rubric-based process for the following three years, allowing us to compare which parts of the applications were used to drive admissions decisions. We find that faculty focused on applicants’ physics GRE scores and grade point averages when making admissions decisions before the implementation of the rubric. While we were able to develop a sufficiently good model whose results we could trust for the data before the implementation of the rubric, we were unable to do so for the data collected after the implementation of the rubric, despite multiple modifications to the algorithms and data such as implementing Tomek Links. Our inability to model the second data set despite being able to model the first combined with model comparison analyses suggests that rubric-based admissions does change the underlying process. These results suggest that rubric-based holistic review is a method that could make the graduate admissions process in physics more equitable.
\end{abstract}

\maketitle

\section{Introduction}
While graduate school has historically been seen as a route for students to begin careers in academia, graduates are increasingly pursuing careers across industry, government, and academia. The National Science Foundation's Survey of Doctorate Recipients finds that less than half of all PhDs work at an educational institution while only 2 out of 5 physics PhDs do \cite{national_center_for_science_and_engineering_statistics_ncses_survey_2021}. As such, universities have a duty to ensure that their students are able to achieve their chosen career trajectories.

Yet, the data suggests that isn't always the case. Only 3 out of 5 physics students who enroll in a PhD program will successfully complete their program \cite{miller_typical_2019,king_ph_2008}. As undertaking graduate study involves a significant time and financial investment from both the student and institution, failing to ensure students graduate leads to a waste of resources. Solutions must consider both the admission and retention sides to this problem. In this paper, we will focus on the former.

As the Council of Graduate Schools notes in one of its reports, ``Better selection [of graduate students] can result in higher completion rates" \cite{council_of_graduate_schools_ph_2004}. Historically and continuing to today, graduate school admissions in the US have tended to be an exclusionary process that favors certain groups over others. Previous research into the graduate admissions process in physics has found that the process relies heavily on the quantitative metrics such as grade point average (GPA) and General and Physics GRE scores \cite{potvin_investigating_2017,chari_admissions_2019,chari_understanding_2019,posselt_inside_2016,posselt_metrics_2019,doyle_search_2015}. These metrics have been found to benefit groups already overrepresented in higher education. For example, prior work has shown students from groups underrepresented in higher education (e.g., first generation, low income, Black, Latinx, Native) suffered a grade penalty relative to their more privileged peers with students from minoritized racial groups suffering the largest penalties \cite{whitcomb_not_2021}. Other work has shown that the General and Physics GREs are biased against women and students from minoritized racial and ethnic groups \cite{miller_test_2014,miller_typical_2019} as well as against students from smaller or less prestigious universities \cite{mikkelsen_investigating_2021}. Furthermore, the high costs associated with these often-required tests, despite limited evidence that these tests serve a predictive purpose \cite{verostek_analyzing_2021,miller_typical_2019,wilkerson_relationship_2007}, prevent some students from applying \cite{cochran_identifying_2018,owens_physics_2020}.

The inequities in the admissions process and the fact that traditional admissions methods ``miss many talented students" \cite{rudolph_phd_2019} have led various programs and organizations to consider alternative admission approaches such as holistic admissions, which considers a ``broad range of candidate qualities including ‘noncognitive’ or personal attributes” \cite{kent_holistic_2016}. These efforts are often supported by rubrics to ensure that all applicants are assessed on the same explicit criteria, provide a structure to do so, and reduce implicit bias present in the admissions process \cite{posselt_inside_2016,miller_equitable_2020}.

Initial results from physics and related disciplines such as engineering suggest that rubric-based holistic admissions can achieve these goals and may lead to increased rates of admission for women and students of minoritized races and ethnicities \cite{stiner-jones_2019_2020,young_physics_2021,barker_work_2021}. However, it is difficult to know whether the rubrics are changing the underlying inequitable admissions process currently in use or if they are merely new tools to perpetuate the same inequities. For example, even in departments actively working to increase their diversity,  prior work has found that GPA and GRE scores had an undue influence on who was admitted \cite{posselt_metrics_2019}. 

Therefore, thinking about how to address inequities in graduate admissions in physics, we ask: how does the introduction of a rubric change a program's admissions process? Framing the question as a computational modeling problem, we operationalize this question into two research questions, using our department's graduate admissions process as an example:

\begin{enumerate}
    \item How do admissions models before and after the implementation of the rubric compare in terms of predictive ability and meaningful features when our models are based on the data contained in applications?
    \item How does using the data produced by faculty when rating applicants using the rubric affect our ability to create admissions models?
\end{enumerate}

To answer these questions, we compare admission models of the current process using data from both faculty ratings and the applications to historical data of the program’s initial process. In our initial analysis of the historical data \cite{young_using_2020}, we noticed there are cases where applicants have similar physics GRE scores and GPA, yet one applicant is accepted while the other is not. Given that cases such as these might add challenges to modeling the data, removing such applicants might allow us to better characterize the general trends in the data. We therefore consider a new approach that detects similar applicants with different admission outcomes and removes them from the data set: Tomek Links \cite{ivan_tomek_two_1976}. We then ask a third research question:

\begin{enumerate}[resume]
    \item How does using Tomek Links affect our ability to model the admissions data, both before and after the implementation of the rubric?
\end{enumerate}

Unlike other studies in physics graduate admissions, this work represents a case study of a single institution rather than a broad look at the graduate admissions landscape. However, because physics is regarded as a high consensus discipline, that is, there is large agreement about what counts as legitimate admissions practices \cite{posselt_disciplinary_2015}, we believe our results will be relevant to similar doctoral programs.

\section{Framing admissions as a computational problem}

\begin{figure*}
    \centering
    \includegraphics[width=.95\linewidth]{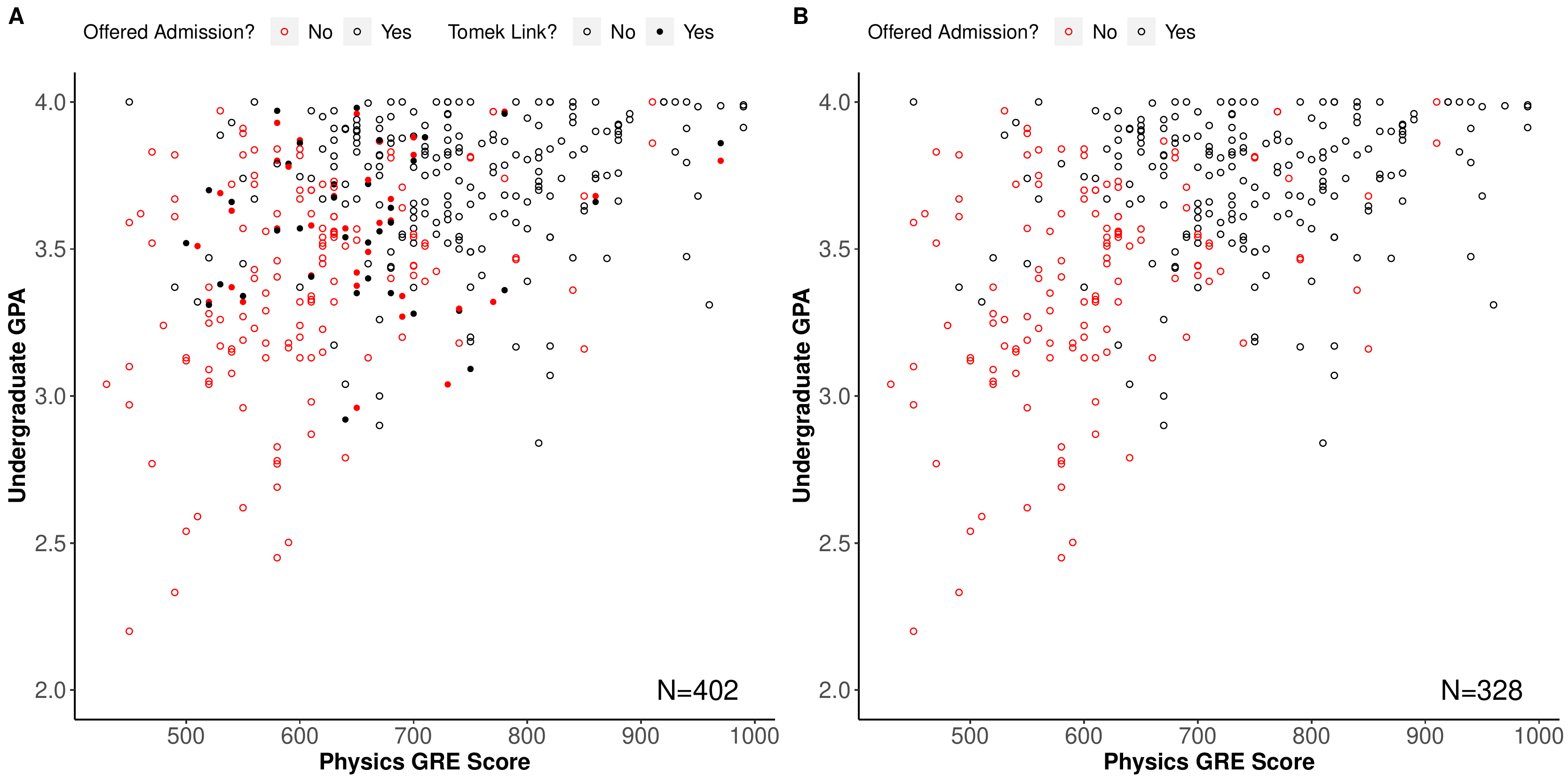}
    \caption{Plot A shows Figure 2 of Young and Caballero \cite{young_using_2020} with the Tomek Links marked. Filled points represent Tomek Links. Plot B shows the same plot after the Tomek Links have been removed. Data for which either the GPA or physics GRE score is missing is not plotted.}
    \label{fig:tomek_example}
\end{figure*}

When evaluating applicants to a graduate program, faculty are presented with information about the applicant and must make a judgement as to whether to admit or reject the applicant. Whether the applicant is admitted or rejected depends on a set of criteria developed by the faculty members reviewing the applicant. As such, we choose to frame the problem of understanding admissions as a classification problem, where a computer must use a set of rules to determine what the qualitative outcome should be or was \cite{hastie_elements_2009}.

Here, we will make the assumption that faculty are primarily seeking to admit applicants who are likely to succeed in their graduate program. As Small notes, there are other possibilities such as aligning with research needs or funding, increasing diversity and inclusion, or developing talent from a specific geographic area \cite{small_range_2017}. We will also assume that these applicants are included in the data but represent only a small fraction of the cases.

When evaluating the potential ``success" of an applicant in the program, there will likely be cases where an argument for and against admission can be made. While admissions committees use common criteria for initially judging applicants, deliberations of these borderline applicants under the traditional process might come down to subtle distinctions that were not used for other applicants \cite{posselt_inside_2016}. Thinking in terms of a modeling perspective, this means that some applicants might be assessed according to additional and potentially implicit criteria and hence, these borderline applicants might not be easily classified by a general model of the admissions process. As a result, including these borderline applicants might cause our model performance to suffer. Alternatively, excluding these applicants and instead focusing on a more typical applicant could improve model performance and provide better insight into whether the underlying process changed.

Unfortunately, whether an applicant is a borderline applicant is not included in faculty ratings of applicants and hence, we do not know who is a borderline applicant. To determine who might be a borderline applicant, let us assume there is a predictive model of a graduate admissions process that perfectly separates those who are admitted and those who are not admitted in some $n$-dimensional application space. We could then say that those applicants who are near the $n-1$-dimensional boundary that separates the admitted applicants and not admitted applicants are borderline applicants. To differentiate borderline applicants in the admissions process from borderline applicants in the modeling process, we will refer to the latter as \textit{boundary applicants}. Such a definition of \textit{boundary applicants} is similar to Hoens and Chawla's definition of borderline cases in classification, which are cases where a small change in the features would cause the classification boundary to shift \cite{hoens_imbalanced_2013}.

However, such an approach assumes that those who are admitted and not admitted can be cleanly split in some $n$-dimensional space and are not intermixed. For a variety of reasons (such as those listed in Small \cite{small_range_2017}), an applicant with a stellar application might be rejected or an applicant with a weaker application might be admitted and hence an admitted applicant might fall on the not-admitted side of the separating boundary or vice versa. While these applicants might not be borderline in the traditional sense, their admission decision likely would have required deliberation and hence, might have gone through a similar process as a borderline applicant. We should therefore also consider these applicants as borderline applicants in the sense of the possibility of hurting our model's performance. Perhaps more accurately, we should refer to these applicants as \textit{noise applicants} following Hoens and Chawla's definition of noise cases, which are cases that result from random variation and are not representative of the underlying pattern \cite{hoens_imbalanced_2013}.

While we have operationalized borderline applicants in terms of a model as \textit{boundary applicants} and \textit{noise applicants}, we still need a method to determine which applicants these are before constructing any models. Tomek Links offers one possible method as it is a method of identifying the boundary or noise cases in the data \cite{ivan_tomek_two_1976}.

To identify the Tomek Links in a data set, the distances between all cases in the data set are computed. Using the distances, the nearest neighbor of each case is computed. For two cases, e.g. case 1 and case 2, the cases are Tomek Links if and only if case 1 is the nearest neighbor of case 2, case 2 is the nearest neighbor of case 1, case 1 and case 2 are of different classes. The only way for these conditions to be fulfilled is if case 1 and case 2 are boundary cases or if case 1 or case 2 is a noise case \cite{hoens_imbalanced_2013}. Therefore, Tomek Links allows us to identify \textit{boundary applicants} and \textit{noise applicants} in our data. An example of this approach in practice is shown in Fig. \ref{fig:tomek_example}.

While Tomek Links have been successfully used in other contexts (e.g. see \cite{min_zeng_effective_2016,batista_study_2004,sawangarreerak_random_2020}), these approaches have tended to use data augmentation in conjunction with Tomek Links. While data augmentation approaches are valid from a modeling perspective, they might be questionable from an ethics and policy perspective. For example, altering the data set might lead to a model that is highly inaccurate of the underlying process \cite{sharma_data_2020}. For our data set, using data augmentation is analogous to creating applicants and thus our conclusions about how our admissions process might or might not have changed would be based on both real and imaginary applicants. For this reason, we will not use data augmentation.

As we note in our methods, we do impute our data. Readers may view this as a contradiction of the previous paragraph but we view data imputation and data augmentation as different. Data imputation is using the existing data to fill in the missing values. In the case of multiple imputations \cite{rubin_basic_1986,van_buuren_flexible_2018}, which we use in this study, the filling in happens multiple times in multiple ways so that the results represent the average result across many possible ways the complete data set might have looked. In contrast, data augmentation is using the existing data to create new data rather than fill in ``holes'' in the data. More generally, data imputation is estimating the results as if we knew the values of the missing data while data augmentation is creating new data to simulate a bigger data set.

\begin{table*}[!htb]
\centering
\caption{The three models compared in this paper and the data that went into each}
\label{tab:admissions_models}
\begin{tabular}{p{0.1\linewidth}p{0.4\linewidth}p{0.09\linewidth}p{0.1\linewidth}p{0.15\linewidth}} \hline
Name & Data source and features & Number of Domestic Applicants & Percent Admitted & Where results are reported \\ \hline
Data Set 0 & Information pulled from the applications before our department implemented a rubric (2014-2017). Features are shown in Table \ref{tab:pre_factors}  & 512 & 48\% & Section \ref{sec:pre_results}\\
Data Set 1a & Information pulled from the \textit{applications} after our department implemented a  rubric (2018-2020). Features are shown in Table \ref{tab:pre_factors} & 511 & 34\% & Section \ref{sec:1a_results}\\
Data Set 1b & Rubric ratings generated \textit{by} faculty as they evaluated applications (2018-2020).Features are shown in the appendix of \cite{young_rubric-based_2021}. & 321 & 43\% & Section \ref{sec:1b_results}\\ \hline
\end{tabular}
\end{table*}

\section{Methods}
In this section, we describe how we collected and processed the data, how we converted undergraduate institutions into data meaningful to our model, the algorithms we used, and how we implemented them.

\subsection{Preparation}
Data for this study comes from applications to the physics graduate program at Michigan State University to enroll in fall 2014 through fall 2020. The admissions process is unique at this university in that the applications are not only reviewed by a central committee but also members of the subdisciplines in which the student expresses interest. Domestic and international applicants do not undergo the same review process and hence we only analyze applications from domestic students. Here, a domestic student is defined to be a U.S. citizen or permanent resident. 

Applicants submitted general and physics GRE scores, transcripts, a personal statement, a research statement, and letters of recommendation. Per a ballot initiative in the state of Michigan, Michigan State University and the other Michigan public universities are explicitly prohibited from discriminating against or granting preferential treatment to individuals based on race, sex, color, ethnicity, or national origin in education \cite{noauthor_constitution_nodate}. To comply with this law, our university's admissions system collects limited demographic data and our department chose not to record the information that was available when evaluating applicants.

Data from applicants planning to enroll between 2014-2017 was obtained through departmental spreadsheets that recorded key information from the applicants as compiled by the admissions chair. This data included general and physics GRE scores, GPA, research subfield of interest, and undergraduate institution.

Starting with the cohort to begin our program in fall 2018, the admissions committee used a rubric to rate applicants on 18 criteria, covering academic preparation, research, non-cognitive competencies/personality traits, fit with the program, and GRE scores. We also obtained these ratings as compiled by the graduate chair. More details about the process and the rubric can be found in Young et al \cite{young_physics_2021}.

In addition, we manually went through the applications for this cohort to extract the same information as was available for the cohort planning to enroll between 2014-2017 to form a comparative data set. Details of the process and data handling are also described in Young et al \cite{young_physics_2021}. Applications from the 2014-2017 cohort were not available to us and hence, we could not perform the same process for that cohort.

For convention, we will refer to data collected before the implementation of the rubric (fall 2014 - fall 2017) as \textit{data set 0} following the convention of using ``naught'' for initial time in physics and data collected after the implementation of the rubric (fall 2018 - fall 2020) as \textit{data set 1}, following the convention of using ``1'' to be mean the next time the data was collected. Furthermore, data in data set 1 that comes from the applications will be referred to as the \textit{data set 1a} while data that comes from the faculty ratings using the rubric will be referred to as \textit{data set 1b} data. These are summarized in Table \ref{tab:admissions_models}

\subsubsection{Describing Undergraduate Institutions}
Because the name of the undergraduate institution in itself does not provide useful information to an algorithm, we created new factors to describe characteristics of the institutions. To describe the overall institution, we classified each institution as public or private, whether it is a minority serving institution (MSI), the region of the country it is located in (such as Northeast, Southwest, etc.), and the Barron's selectivity of the institution, which describes how selective the undergraduate program is. We assume that selectivity serves as a proxy for prestige. Classifications for the first three categories were taken from the most recent Carnegie Rankings \cite{indiana_university_center_for_postsecondary_research_carnegie_nodate} while the Barron's classification came from Barron's \textit{Profiles of American Colleges} \cite{national_center_for_education_statistics_nces-barrons_2017}. Because the overall reputation of the applicant's undergraduate university might not describe the physics program at that university, we also included factors related to the physics program, such as the highest physics degree offered at the university and the size of the undergraduate program and PhD program if applicable. The size of the undergraduate and PhD programs were determined by the median number of graduates of the program between the 2012-2013 and 2015-2016 academic years for data set 0 and 2016-2017 through 2018-2019 for data set 1a (i.e. the years that applicants applied to the program). The programs were then classified as small, medium-small, medium-large, or large based on which quartile they fell into.  We used the Roster of Physics Departments with Enrollment and Degree Data to collect this data \cite{nicholson_roster_2014,nicholson_roster_2015,nicholson_roster_2016,nicholson_roster_2017,nicholson_roster_2018,nicholson_roster_2019,nicholson_roster_2020}. All factors used in the models for data sets 0 and 1a are shown in Table \ref{tab:pre_factors} and include the scale of measurement.

\begin{table}[]
\centering
\caption{Features used in our model of data sets 0 and 1a, including their scale of measurement}
\begin{tabular}{p{0.55\linewidth}p{0.35\linewidth}}
\hline \textbf{Feature} & \textbf{Measurement Scale} \\ \hline
Undergraduate GPA    & Continuous        \\
Verbal GRE score    & Continuous        \\
Quantitative GRE score   & Continuous        \\
Written GRE score      & Continuous        \\
Physics GRE score     & Continuous        \\
Proposed research area & Categorical       \\
Application year    & Categorical       \\
Barron's selectivity    & Categorical       \\
Region of applicant's undergraduate institution  & Categorical       \\
Type of physics program at applicant's undergraduate institution    & Categorical       \\
Size of undergraduate physics program at applicant's undergraduate institution & Categorical       \\
Size of doctoral physics program at applicant's undergraduate institution      & Categorical       \\
Applicant attended a minority serving institution                    & Binary            \\
Public or Private  & Binary \\ 
\\
Output variable: admitted status & Binary \\
\hline
\hline
\end{tabular}
\label{tab:pre_factors}
\end{table}

\subsubsection{Justifying our choice of institutional factors}
% Source for historical News & World Rankings: https://andyreiter.com/datasets/
Prior work has documented university pedigree is often considered in the application process because institutional quality is assumed to be a proxy for student quality \cite{posselt_inside_2016, paxton_perceived_2003}. Here, we measure institutional quality by Barron's selectivity and public or private status, with the assumption that physics faculty view private universities as more prestigious than public universities. For example, US News \& World, publisher of a well-known college ranking system, has not included a public university in its top 10 in the past decade and no more than 1 public university in its top 20. We include also region of the applicant's undergraduate university to account for the fact that the institution being studied is a public university and might therefore show a preference for students from the surrounding region.

Prior work has also found faculty exhibit a tendency to admit students like themselves, though it is more common among academics who graduated from elite institutions \cite{posselt_inside_2016}. Therefore, it is not unreasonable to expect that faculty may prefer to admit students who followed similar paths as they did, meaning students from large, doctoral institutions might be more likely to be admitted than students from smaller institutions. Additionally, we use the size of the undergraduate and PhD programs as proxies for the perceived prestige of the physics department, assuming a more prestigious physics department attracts more students and hence graduates more students.

\subsection{Analysis}
\subsubsection{The Random Forest Algorithm}
To analyze our data, we used the conditional inference forest algorithm, a variant of the random forest algorithm \cite{breiman_random_2001} shown to be less biased when the data includes both continuous and categorical variables \cite{strobl_bias_2007} such as those used in our model (see Table \ref{tab:pre_factors}). Random forest models in general are ensembles of individual decision trees, which use binary splits of the input features to make a prediction. The predictions are then averaged and sometimes weighted over the individual trees to obtain the overall prediction of the random forest. 

While there are multiple metrics used to assess random forest and other machine learning models, two of the most common are the accuracy and the area under the curve (AUC). The accuracy is simply the proportion of correct predictions made by the model. To ensure that the accuracy is not inflated by overtraining, only a fraction of the available data is used to construct the model while the rest is used to test the predictive power. It is this remaining data that is used to calculate the accuracy of the model. This process of splitting data into training and testing sets is often repeated multiple times to understand the variation in the accuracy or other metric of the model through a process called \textit{cross validation (cv)}.

The AUC is defined as the area beneath the receiver operator curve of the model, which visualizes the false positive rate against the true positive rate and varies between 0.5 and 1, with values greater than 0.7 signifying an acceptable model \cite{araujo_validation_2005}. The area describes the proportion of positive cases that are ranked above negative cases in the data set by the model. For example, for our data, the AUC would represent the proportion of all random pairs of admitted and not-admitted applicants in which the admitted applicant is classified as admitted and the not-admitted applicant is classified as not-admitted.

In addition to making predictions, the random forest algorithm can determine the importance of each feature to the model, referred to as the feature importance. For this analysis, we use two importance measures. First, we used the AUC permutation feature importance \cite{janitza_auc-based_2013} as it is claimed to be less biased than the accuracy-based permutation importance when input features differ in scale (as do our factors listed in Table \ref{tab:pre_factors}) and when the predicted variable is not split evenly between the two outcomes. In practice, our previous work suggests which method we pick will have minimal effect on the conclusions \cite{young_predictive_2021}. Under this approach, each feature is randomly permuted and then passed through the model to make a prediction. The AUC is then recorded and the difference between this value and the original AUC is computed. As permuting a feature with more predictive information should result is a worse model than permuting a feature with less predictive information, a larger difference between the original AUC and the AUC with a permuted feature suggests that this feature contains more predictive information. These differences can then be used to create a relative ordering of features.

However, if the features are correlated, it is possible that the orderings may be biased or that permutations of one feature might result in unrealistic combinations of features and hence would cause the model to extrapolate performance \cite{hooker_unrestricted_2021}. For example, if all students who earned perfect scores on the physics GRE also had high GPAs, permuting GPA could cause there to be cases where a perfect physics GRE score goes with a low GPA, which would be outside of the region learned by the model. To prevent that, a conditional importance measure has been proposed in which features are permuted within a subset of similar cases \cite{strobl_conditional_2008}. Because of the correlations between various sections of the GRE (e.g., Verostek et al. reported a moderate correlation between the physics GRE score and the quantitative GRE score \cite{verostek_analyzing_2021}), we also used this conditional approach to compute feature importances.

Feature importances are derived from the data and hence, are not assumed to follow any statistical distribution. Therefore, there is no simple way to apply the idea of statistical significance to feature importances. We instead applied the recursive backward elimination technique described in D\'{i}az-Uriarte and Alvarez de Andr\'{e}s \cite{diaz-uriarte_gene_2006} to determine which features are predictive of admission and which are not. When using this technique, the features are ordered according to their importance. A model is then built using all the features and the accuracy is computed. A set fraction of the features with the smallest importances are then removed and a new model is built and the accuracy computed. This process continues until only 2 features are left. The model with the fewest number of features while maintaining an accuracy within a standard error of the highest accuracy across all models built in this process is then the selected model. We will refer to the features used in this selected model as the \emph{meaningful} features and interpret them as the features that are predictive of the outcome. For more information about random forest models, biases, and feature importance measures, see the supplemental material of Young et al. \cite{young_identifying_2019}.

We chose to apply a random forest model instead of a more traditional technique for classifying data such as logistic regression (as used by Attiyeh and Attiyeh \cite{attiyeh_testing_1997} and Posselt et al. \cite{posselt_metrics_2019} to study graduate admissions) due to these feature importances. As feature importances measure all factors on the same scale, that is, how much they change the area under the curve, factors of otherwise different scales can be compared. This contrasts with logistic regression where the odds ratio for a continuous variable would measure the change in odds for a unit increase in the variable while the odds ratio for a categorical or binary variable measures the change in odds relative to a reference group. In addition, the feature importances allow for each categorical feature as a whole to be compared to the other features rather than in pairs relative to the reference group. This property can be especially useful for features like proposed research area where there is no natural or standard choice of reference group or when we are not interested in category differences.

\subsubsection{Comparing different classification models}
When using multiple classification models on a data set, an important consideration is how to compare the different models and determine the best one. Simple methods to do so include comparing a metric of interest such as the accuracy or the AUC and choosing the model with the highest average value over the data sets or picking the model that has the highest metric on the largest number of data sets. \cite{demsar_statistical_2006}.

However, it is also possible that one model may appear to be better than another due to chance. Therefore, a test of statistical significance may be of interest to better understand whether that might be the case. Dietterich compared five such methods for doing so and Alpaydin developed a more robust version of the 5x2 cv paired t-test method preferred by Dietterich \cite{dietterich_approximate_1998,alpaydm_combined_1999}. We describe Alpaydin's 5x2 cv combined F test below.

Assume that there are two classifiers $A$ and $B$ and a data set $D$. Split $D$ randomly in half, forming a training set and a testing set. Then use the training set to build a model with classifiers $A$ and $B$ and apply those models to the testing set to obtain accuracies $p^{(1)}_A$ and $p^{(1)}_B$. Next, swap the training and testing sets and repeat the procedure, computing testing accuracies $p^{(2)}_A$ and $p^{(2)}_B$. Following this, the difference in testing accuracies between model $A$ and $B$ are computed, $p^{(1)}$ and $p^{(2)}$. Finally, the mean and variance of the differences are computed.

This procedure is then repeated five times. The F-static proposed by Alpaydin is then

\begin{equation}
  f=\frac{\sum_{i=1}^5\sum_{j=1}^2 (p_i^j)^2}{2\sum_{i=1}^5 s_i^2} 
\end{equation}

where $p_i^j$ is the difference in accuracies for the $j$th trial of the $i$th iteration. This $f$ is then approximately distributed as an F-statistic with 10 and 5 degrees of freedom. The $f$ and the degrees of freedom can then be used to calculate the probability of obtaining results given that there is no difference between classifiers $A$ and $B$ or the $p$-value. If the $p$-value is less than some cutoff, $\alpha$, then the classifiers are said to be statistically different. See Alpaydin for details \cite{alpaydm_combined_1999}.

\subsubsection{Implementation}
The implementation of the analysis largely follows the framework detailed in Aiken et al. \cite{aiken_framework_2021}. 

To perform the analysis, we used R \cite{r_core_team_r:_2018} and the \texttt{party} package \cite{hothorn_survival_2006,strobl_bias_2007,strobl_conditional_2008} to create a conditional inference forest model. We used 70\% of our data to train the model, 500 trees to build our forest and used $\sqrt{p}$ as the number of randomly selected features to use to build each tree, with $p$ being the total number of features in the model. These values follow recommendations of Svetnik et al. \cite{svetnik_random_2003}. We ran our model 30 times, randomly selecting 70\% of our data for training each time. For each trial, we calculated the training AUC, testing AUC, testing accuracy, null accuracy, and the permutation AUC importances. We then averaged the results. As the conditional inference forest algorithm has routines built in to handle missing data \cite{hapfelmeier_new_2014}, applicants with missing information were not removed from the data set. However, the conditional importance approach requires there to be no missing values so we used the MICE algorithm \cite{buuren_mice_2011} to fill impute missing data in that case, following Nissen et al.'s recommendation for PER \cite{nissen_missing_2019}. The imputation results were pooled using Rubin's Rules \cite{rubin_basic_1986}.

For data sets 0 and 1a, the same features were used as in Table \ref{tab:pre_factors}, with the size of the physics program factors updated with new data for the post-data models. For data set 1b, all features were treated as categorical (0, 1, or 2) and as in our previous work \cite{young_rubric-based_2021}, any values between a rubric level were rounded up.

In addition, to determine if our models depended on our choice of hyperparameters, we varied the fraction of data to train the model, the number of trees in the forest, and the number of randomly selected features to use to build each tree. We set the training fraction to be either 0.5, 0.6, 0.7, 0.8, or 0.9, the number of trees in the forest to be 50, 100, 500, 1000, or 5000, and the number of features used for each tree to be 1, $\sqrt{p}$, $p/3$, $p/2$, or $p$ for a total of 125 possible combinations (124 new and the original model). These choices are based off findings in Svetnik et al. \cite{svetnik_random_2003}: namely, that the error rates level off once the number of trees is on the order of $10^2$ and their choices of the number of features in each tree. In addition, increasing the training fraction may improve performance as there is more data for the model to learn from. For each combination, we repeated the procedure in the previous paragraph. Due to the computational cost of the conditional permutation approach, we only calculated the AUC-permutation importance.

To determine if changing the hyperparameters affected our models, we computed the minimum, median, and maximum value of each metric over the 125 hyperparameter combinations and the relative ordering of the features in each model. We chose the minimum, median, and maximum instead of the mean and standard error because (1) we are looking across different models rather than getting repeated measurements of the same things so we cannot assume the results will be normally distributed and (2) we are interested in the best and worst performance achieved under hyperparameter tuning to get a sense of the possible values we can achieve which would not be possible using the mean and standard error. If our model is largely unaffected by the choice of hyperparameters, we would expect the metrics to show minimal variation and the relative ordering of the features to be largely unchanged.

To compute the Tomek Links, we used the \texttt{TomekClassif} function in the \texttt{UBL} package \cite{branco_ubl_2016}. We first used MICE to impute the data before calculating the Tomek Links using the function defaults with the exception of the distance metrics. Following the recommendation of the package's documentation, we used the \texttt{HVDM} distance for data sets 0 and 1a because those data sets contain both categorical and continuous data and we used the \texttt{Overlap} distance for data set 1b because all features were categorical.

After removing the Tomek Links, we ran each model 30 times and averaged results. Results were then pooled using Rubin's Rules.

In addition to looking at the feature order to determine if the admission process changed, we can compare the performance of the models themselves. If the process did not change, then a model built from data set 0 should perform equally well (within error) on a data set 0 testing set as on data set 1, and a model built from data set 1a should perform equally well (within error) on a data set 1a testing set as on data set 0. If the process did change, we would expect better performance on the test data pulled from the train/test split than the other data set. In this approach, we are using model fit as a proxy for whether the process changed.

To test this hypothesis, we first randomly split data set 0 into a training and testing set (70\% again to the training set) and built a conditional inference forest model on the training set. We then used the model to predict the testing set and data set 1a, computing the accuracy and AUC. We repeated this process thirty times and averaged the accuracies and AUCs. We then repeated the process for data set 1a by doing a train test split on data set 1a and using all of data set 0 as a testing set. This method provides a simple comparison between the models.

Second, we performed the 5x2 cv combined F-test explained by Alpaydin \cite{alpaydm_combined_1999}. Because our models were not different algorithms, we altered the approach as follows. Both data set 0 and data set 1a were divided into a training and test set, with half of the data in each. For each pair of trials, we used the two training data sets to develop two models (one for the before rubric process and one for after) and applied those models to the testing set from data set 0. We then used the testing set from data set 0 and the testing set from data set 1a to develop two new models and applied those models to original data set 0 training set. The accuracies and AUCs were then subtracted for the same testing set. We then repeated this process five times and computed the $f$ statistic to determine if the models were equally effective at predicting the data before the implementation of the rubric (data set 0).

To determine if the models were equally effective at predicting the data after the implementation of the rubric (data set 1a), we repeated the procedure above, except for swapping the roles of data set 0 and data set 1a.

In both cases, a corrected $p$-value less than $0.05$ would signify a statistically significant difference between the predictive abilities of the models. To correct the $p$-values for multiple comparisons, we use the Holm-Bonferroni procedure \cite{holm_simple_1979}.

\section{Results}\label{sec:results}

\subsection{Understanding the Underlying Data}

\subsubsection{Before Implementation of Rubric (Data Set 0)}\label{sec:pre_results}
\begin{figure}
\centering
  \includegraphics[width=.95\linewidth]{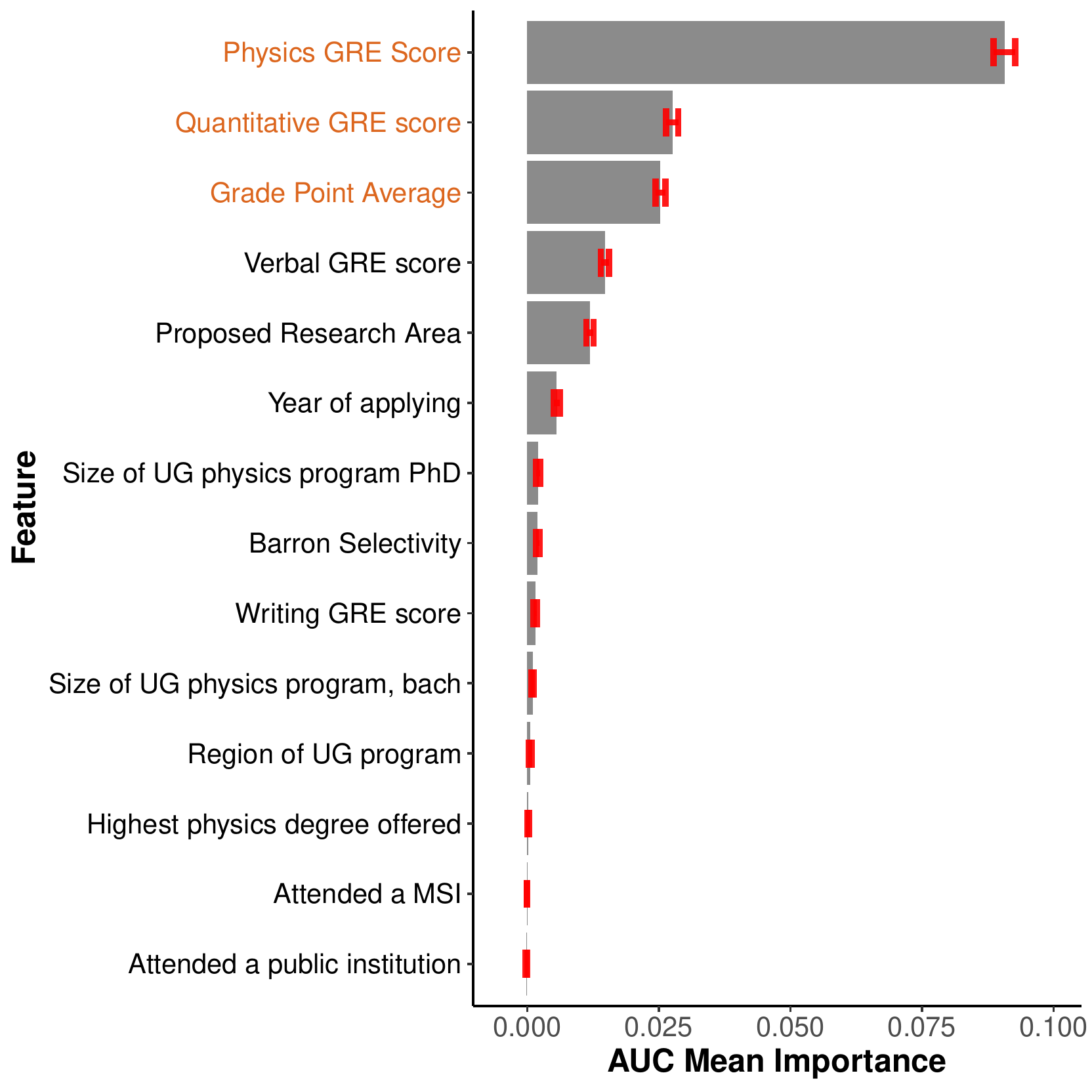}
  \caption{Averaged AUC feature importances over 30 trials for data set 0. Physics GRE score, Quantitative GRE score, and  undergraduate GPA, appearing in orange, were the factors found to be meaningful and hence predictive of being admitted. \label{fig:pre_auc_imp}}
\end{figure}

Across the 30 runs, the average accuracy of our model predicting on the held-out data was $75.6\% \pm 0.6\%$, the average training AUC was $0.849 \pm 0.002$, and the average testing AUC was $0.756 \pm .006$. As our model's accuracy is significantly higher than the null accuracy of $52.7\%$, the percent of students who were not accepted, and our testing AUC is above 0.7, our model can be considered an acceptable model of the data. 

The feature importances averaged over the 30 runs are shown in Fig. \ref{fig:pre_auc_imp}. We find numerical factors such as the applicant's score on the physics GRE, the applicant's score on the quantitative GRE, the applicant's undergraduate GPA, the applicant's verbal GRE score, and their proposed research area to be more important in the application process than any factor describing the applicant's undergraduate institution. Using recursive backward elimination to determine the meaningful factors, we find the applicant's physics GRE score, quantitative GRE score, and their undergraduate GPA to be the only meaningful factors.

To verify that the applicant's physics GRE score, quantitative GRE score, and undergraduate GPA were indeed the only meaningful factors, we then reran our random forest model 30 times using only these three factors as the predictors. Our average testing accuracy was then $75.4\% \pm 0.6\%$ and our testing average area under the curve was $0.754 \pm 0.006$, which are not statistically different from the values we found using all fourteen factors shown in Table \ref{tab:pre_factors}. 

When we instead used MICE and the conditional importances, and the metrics were slightly higher, likely because imputing the missing values provided more data for the algorithm to learn from. Specifically, the testing accuracy was $77.1\% \pm 0.1 \%$ and the testing AUC was $0.770 \pm 0.001$.

\begin{figure}
\centering
  \includegraphics[width=.95\linewidth]{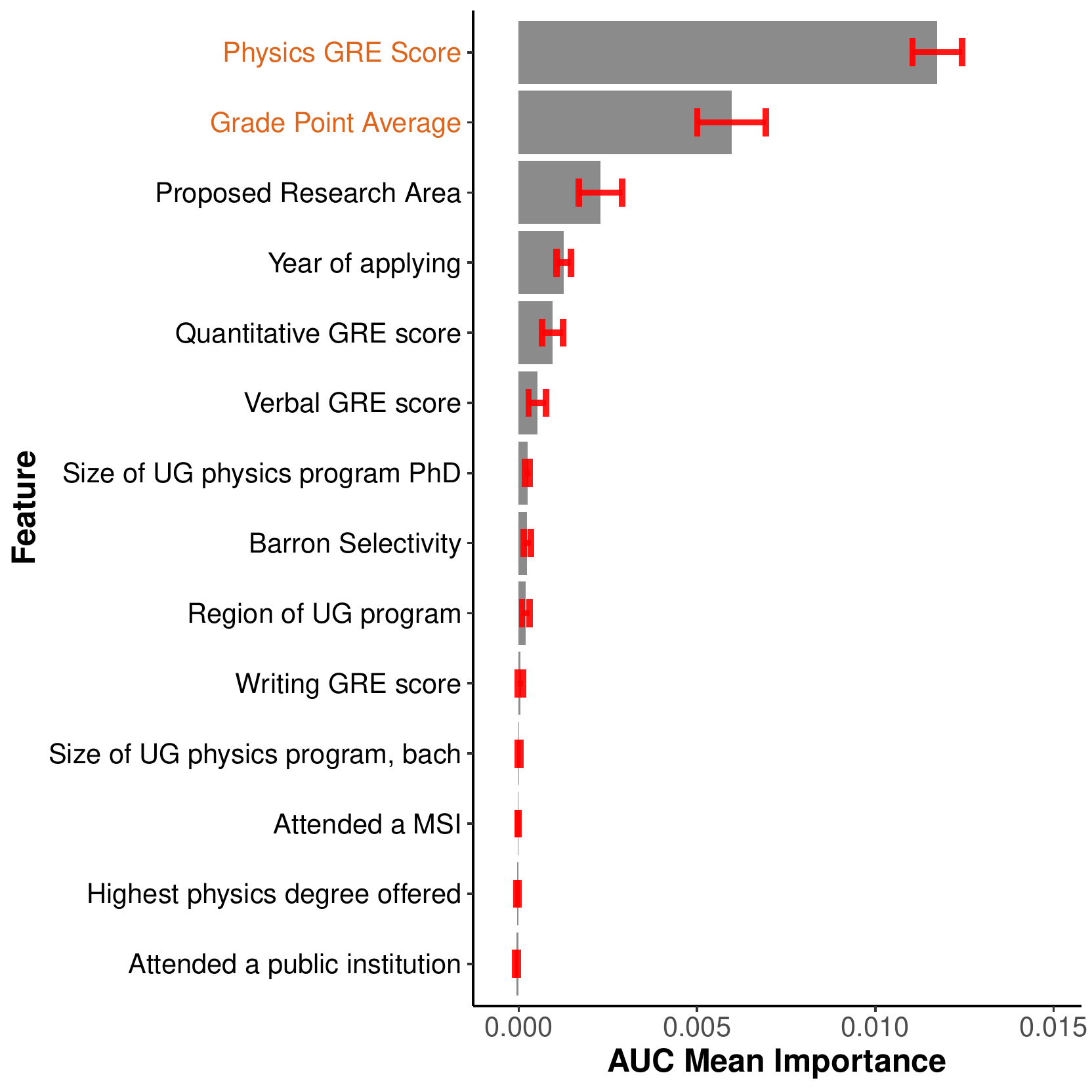}
  \caption{Averaged conditional feature importances over 30 trials for data set 0. Physics GRE score and  undergraduate GPA, appearing in orange, were the factors found to be meaningful and hence predictive of being admitted when adjusting for correlations among the features. \label{fig:pre_cond_imp}}
\end{figure}

The conditional feature importances are shown in Fig. \ref{fig:pre_cond_imp}. Compared to Fig. \ref{fig:pre_auc_imp}, we notice that the verbal and quantitative GRE scores are ranked lower than they were when we did not take correlations into account and proposed research area and year of applying are ranked higher than when we did not take correlations into account. The physics GRE and GPA are still ranked highly however, even after taking correlation into account.

Performing the recursive backward elimination, we find that the physics GRE score and GPA are meaningful features, but the quantitative GRE score no longer is. Using only these two features to create a conditional inference forest on the imputed data, we find that the testing accuracy is $75.7\% \pm 0.7\%$ and the testing AUC is $0.757 \pm 0.007$, which are consistent with the full model.

\begin{table}
\centering
\caption{Minimum, median, and maximum values of the metrics obtained over the 125 hyperparameter combinations.}
\begin{tabular}{lrrr}
  \hline
metric & min & median & max \\ 
  \hline
Train AUC & 0.824 & 0.848 & 0.853 \\ 
  Test AUC & 0.726 & 0.749 & 0.760 \\ 
  Test Accuracy & 0.727 & 0.750 & 0.760 \\ 
  Null Accuracy & 0.521 & 0.527 & 0.556 \\ 
   \hline
\end{tabular}
\label{tab:pre_hyperparameter_metrics}
\end{table}

When we test the various hyperparameter combinations, we find similar results. Looking at the metrics (Table \ref{tab:pre_hyperparameter_metrics}), we see that the testing accuracy varies by 3.3 percentage points between the minimum and maximum values and the testing AUC varies by $0.034$ between the minimum and maximum values. As the variation is limited and these metrics are still within the acceptable range, the results suggest that our choice of hyperparameters has limited impact on the metrics.

When we look at the ranks of the features used in each hyperparameter combination, we also see limited variation. First, we find that physics GRE score, GPA, quantitative and verbal GRE scores, and proposed research area are always the top five features, regardless of the hyperparameters. Second, we find that the institutional features never rank above 7, meaning that no combination of hyperparameters can create a model where these features are predictive of admission. In addition, we notice that year of applying is always ranked sixth, serving as a separating feature from the previous two groups of features. This result is likely due to the fact that there are yearly differences in the fraction of applicants admitted so year is not a noise feature and should be ranked above the noise features. However, knowing the year the applicant applied doesn't say too much about the applicant themselves and hence, we would expect it to rank below the features like test scores and GPA that do.

Looking at the most important features, we notice that physics GRE is always the top ranked feature followed by either GPA or quantitative GRE score, with GPA being the more common selection. Furthermore, GPA never ranks lower than third while the quantitative GRE score ranks between second and fourth. For certain choices of hyperparameters, the applicant's proposed area of research ranks higher than the quantitative GRE score.

For interested readers, the distribution of the ranks is shown in the supplemental material.

\subsubsection{After the Implementation of Rubric: Application Data (Data Set 1a)}\label{sec:1a_results}
Across the 30 runs, the average accuracy of our model predicting on the held-out data was $71.4\% \pm 0.6\%$, the average training AUC was $0.720 \pm 0.004$, and the average testing AUC was $0.626 \pm .006$, which is less than the minimum of 0.7 for a reasonable model. Our null accuracy was $66.0\%$ which suggests that our model is only doing slightly better than if it were to predict everyone was not admitted to our program as the majority of applicants were not admitted.

\begin{figure}
    \centering
    \includegraphics[width=.95\linewidth]{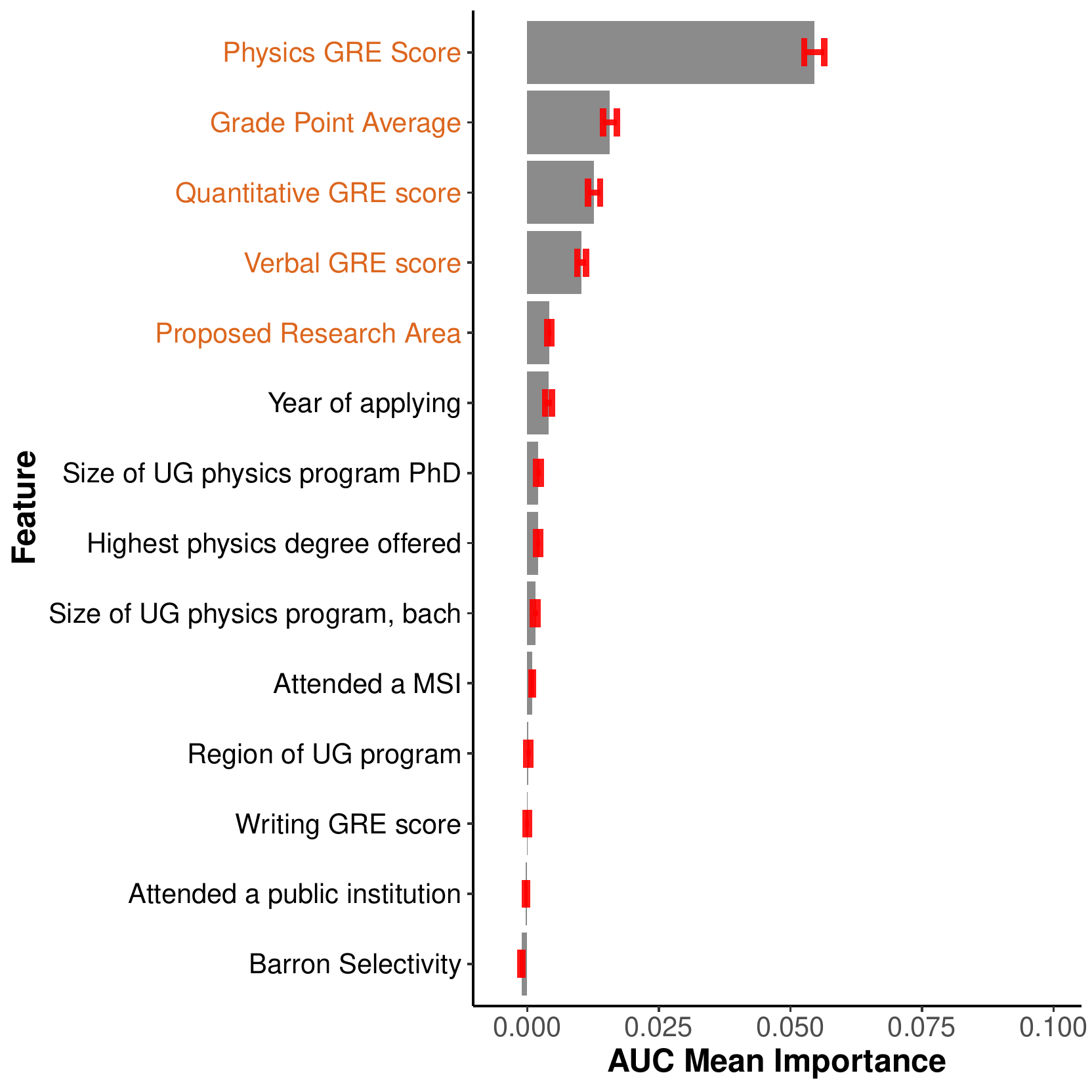}
    \caption{Averaged AUC feature importances over 30 trials for data set 1a. Physics GRE score, undergraduate GPA, Quantitative GRE score, Verbal GRE score and proposed research area, appearing in orange, were the factors found to be meaningful and hence predictive of being admitted.}
    \label{fig:post_auc_imps}
\end{figure}

When looking at the feature importances (Fig. \ref{fig:post_auc_imps}), we see that the physics GRE score, undergraduate GPA, quantitative and verbal GRE scores, and proposed research area are near the top while the institutional features near the bottom. Performing the backward elimination, we find that physics GRE score, undergraduate GPA, quantitative and verbal GRE scores, and proposed research area are the meaningful features predictive of admission after the implementation of the rubric.

When we compare the ranks of the features from data set 1a to the ranks of the features from data set 0, we notice that the order of features is largely unchanged for the most predictive features, with only the quantitative GRE score and GPA switching places. The major difference between the features predictive of admission in data sets 0 and 1a is the number of meaningful features, with data set 0 having three meaningful features and data set 1a having five meaningful features.

\begin{figure}
    \centering
    \includegraphics[width=.95\linewidth]{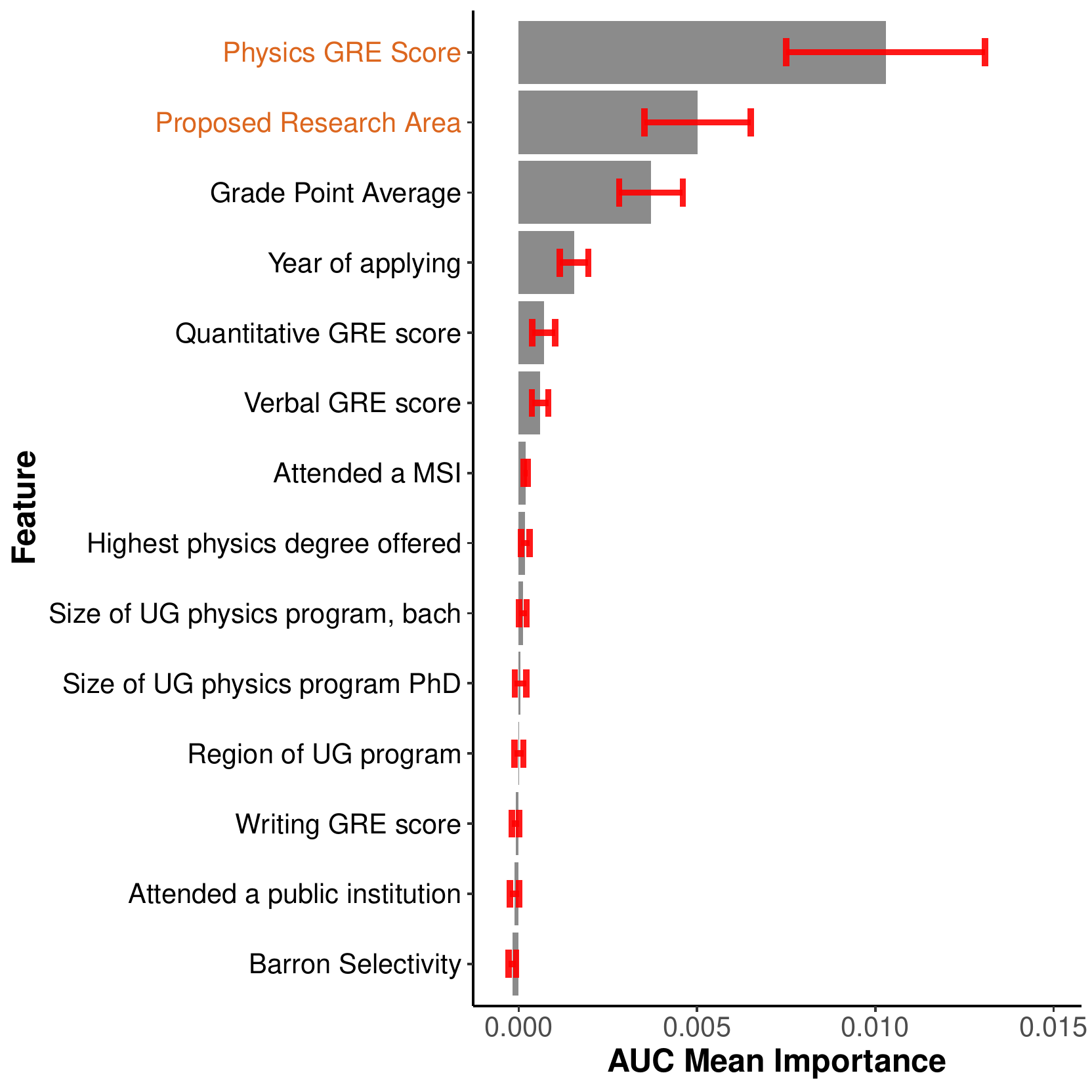}
    \caption{Averaged conditional feature importances over 30 trials for data set 1a. Physics GRE score and proposed research area, appearing in orange, were the factors found to be meaningful and hence predictive of being admitted once correlations were accounted for.}
    \label{fig:post_conditional_imps}
\end{figure}

When we take the correlations among the features into account using the conditional importance measure, however, we notice the set of meaningful features shrinks. As shown in Fig. \ref{fig:post_conditional_imps}, only the applicant's physics GRE score and proposed research area were found to be predictive of admission. It is also important to note that the quantitative and verbal GRE scores are ranked lower once correlations are accounted for, suggesting that their initial importances were inflated.

\begin{table}
\centering
\caption{Minimum, median, and maximum values of the metrics obtained over the 125 hyperparameter combinations for models built from data set 1a.}
\begin{tabular}{lrrr}
  \hline
metric & min & median & max \\ 
  \hline
Train AUC & 0.602 & 0.735 & 0.749 \\ 
  Test AUC & 0.549 & 0.633 & 0.676 \\ 
  Test Accuracy & 0.679 & 0.712 & 0.732 \\ 
  Null Accuracy & 0.645 & 0.661 & 0.666 \\ 
  \hline
\end{tabular}
\label{tab:post_hyperparameter_metrics}
\end{table}

Given the poor performance of our model (AUC $<$0.7, testing accuracy only slightly higher than the null accuracy), hyperparameter tuning might have improved the model. While it did to a degree, the testing accuracy was still only a few percentage points above the null accuracy and the testing AUC was still below 0.7 (Table \ref{tab:post_hyperparameter_metrics}). Thus, even with hyperparameter tuning, the models of data set 1a were poor.

Finally, to see how the feature ranks varied based on the hyperparameters, we plotted the occurrence fraction of each rank for each feature (see appendix). We notice that across the 125 hyperparameter combinations, physics GRE score and GPA are almost always the top two features, followed by quantitative and verbal GRE scores. In addition, none of the institutional features ever rank in the upper half of the importances.

\subsubsection{After the Implementation of Rubric: Rubric Data (Data Set 1b)}\label{sec:1b_results}
Given that after the implementation of the rubric applicants are rated on the rubric constructs, perhaps using the rubric constructs instead of the application data in a model would lead to better performance. It did not. 

We find that the testing AUC was 0.664 $\pm$ 0.007 and the testing accuracy was 0.675 $\pm$ 0.007 (null accuracy 0.553 $\pm$ 0.006). Given that not all applicants had sufficiently complete applications to be reviewed by faculty and those with incomplete applications tended to be not admitted, the null accuracy is smaller for models of data set 1b than the models of data set 1a

\begin{figure}
    \centering
    \includegraphics[width=.95\linewidth]{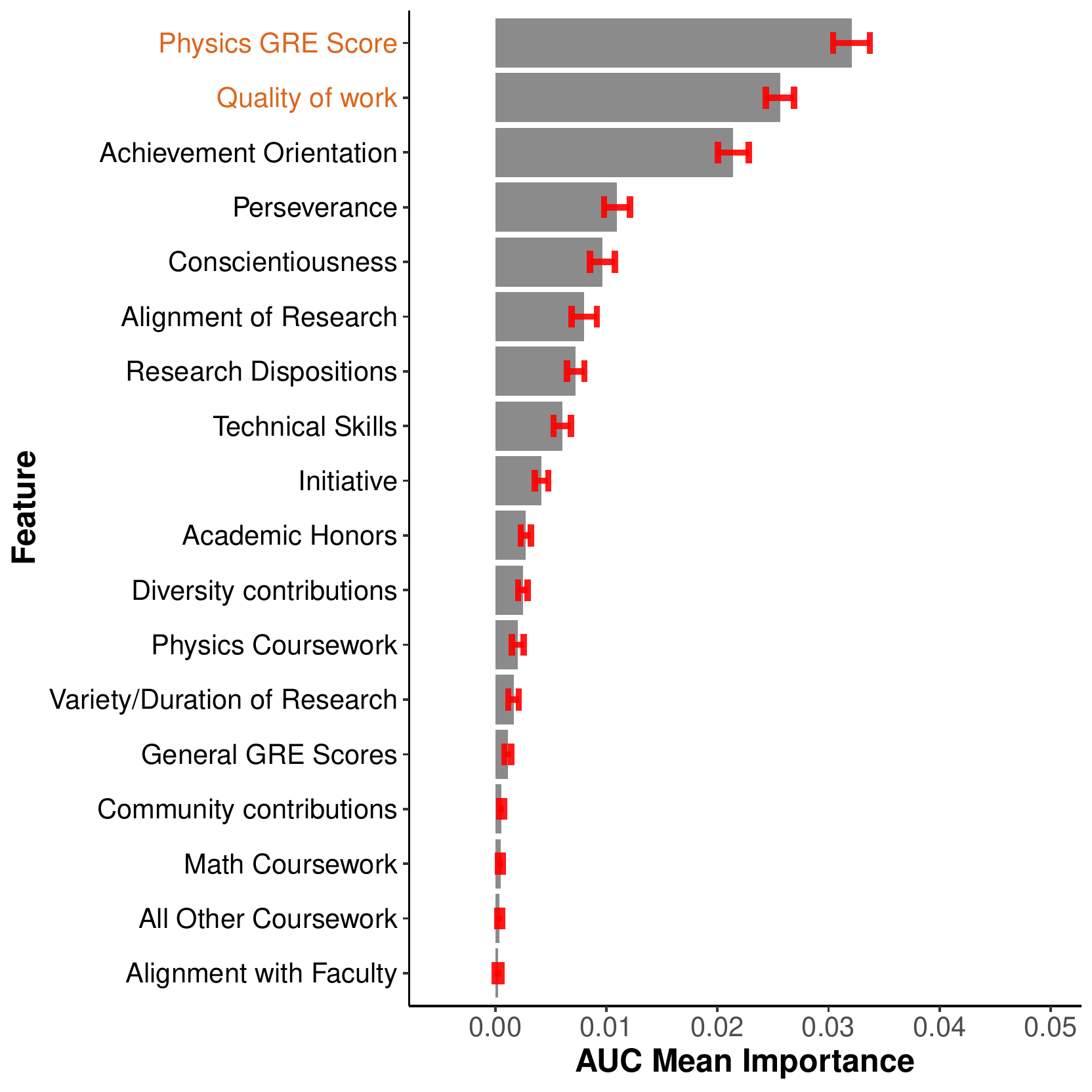}
    \caption{Averaged feature importances over 30 trials for the models of data set 1b. Physics GRE score and quality of work, appearing in orange, were the factors found to be meaningful and hence predictive of being admitted.}
    \label{fig:rubric_auc_imp}
\end{figure}

When we looked at the feature importances, the results showed similarities to the importances from the models of data set 1a. From Fig \ref{fig:rubric_auc_imp}, we notice that physics GRE score is still the top feature. However, measures of GPA such as physics coursework, math coursework, and all other coursework tended to be in the lower half of the rankings, alignment of research (the closest construct to proposed research area) was toward the middle of the rankings, and general GRE scores was toward the bottom despite GPA, proposed research area, and general GRE scores being top ranking features under the models of data set 1a.

From the figure, we also notice that measures related to research (quality of work, research dispositions, and technical skills) are ranked in the upper half as are measures of noncognitive skills (achievement orientation, perseverance, and conscientiousness) while measures of fit (diversity contributions, community contributions, and alignment with faculty) are ranked in the bottom half of features.

When performing the backward elimination, we find that only physics GRE score and quality of work are selected, suggesting that only these two features are needed to produce similar predictive performance as using all 18 features.

\begin{figure}
    \centering
    \includegraphics[width=.95\linewidth]{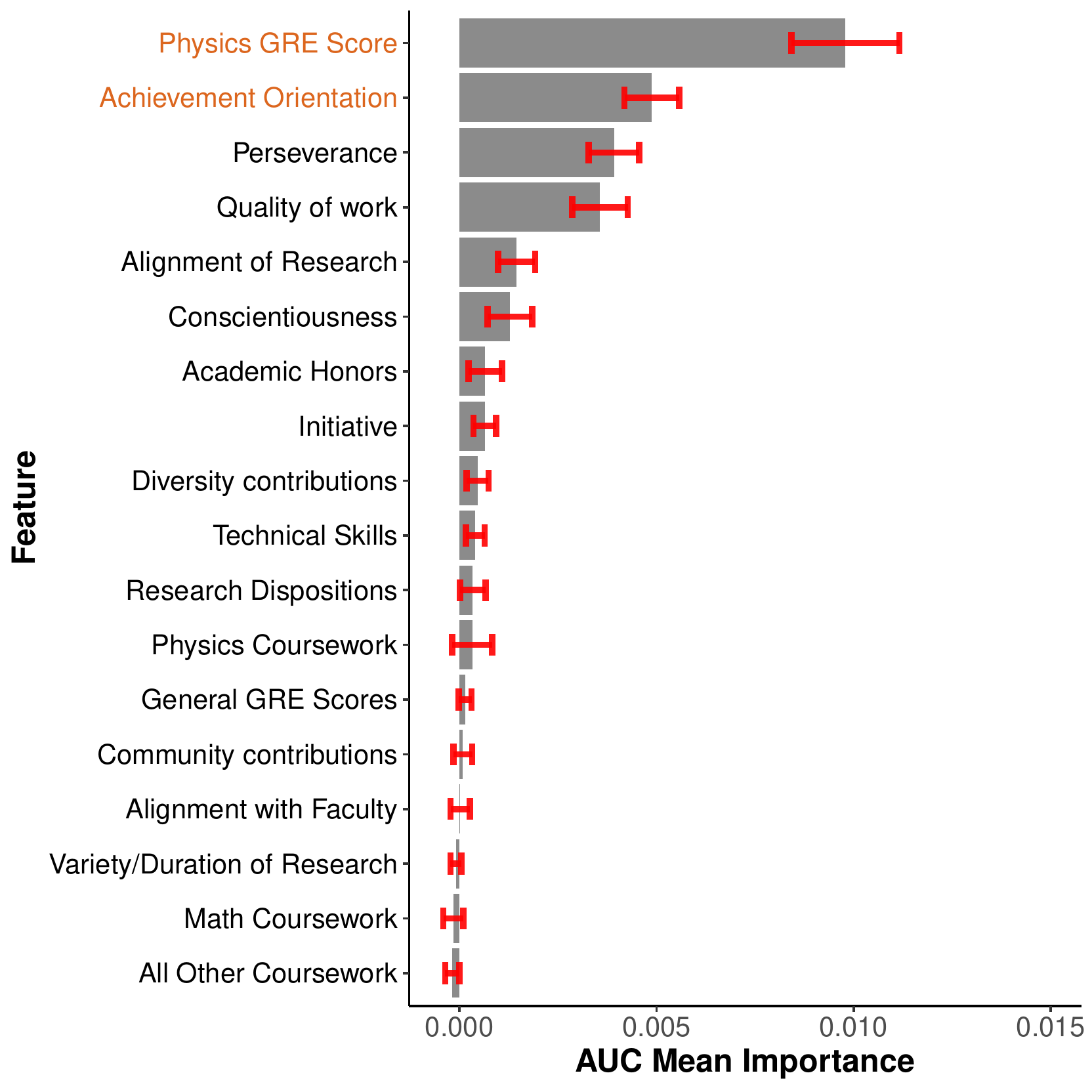}
    \caption{Averaged conditional feature importances over 30 trials for the models of data set 1b. Physics GRE score and achievement orientation, appearing in orange, were the factors found to be meaningful and hence predictive of being admitted once correlations were accounted for.}
    \label{fig:rubric_conditional_imps}
\end{figure}

We then repeated the analysis taking correlations between features into account using the conditional feature importance. The result is shown in Fig. \ref{fig:rubric_conditional_imps}. We notice that the top features are similar, though the rank of quality of work decreased to fourth. Now, physics GRE score and achievement orientation were found to be meaningful and hence predictive features.

\begin{table}
\centering
\caption{Minimum, median, and maximum values of the metrics obtained over the 125 hyperparameter combinations for the models of data set 1b.}
\begin{tabular}{lrrr}
  \hline
metric & min & median & max \\ 
  \hline
Train AUC & 0.711 & 0.767 & 0.791 \\ 
  Test AUC & 0.654 & 0.669 & 0.686 \\ 
  Test Accuracy & 0.660 & 0.678 & 0.696 \\ 
  Null Accuracy & 0.559 & 0.561 & 0.586 \\ 
  \hline
\end{tabular}\label{tab:rubric_hyperparameter_metrics}
\end{table}

Finally, we performed hyperparameter tuning to determine if we could create a model with acceptable metrics. Unfortunately, we could not. Even the best AUC among the 125 hyperparameter tuning combinations did not exceed 0.7. The full results are shown in Table \ref{tab:rubric_hyperparameter_metrics}.

Looking at the feature ranks, we again found a diagonal pattern toward the upper left of the plot, suggesting the same few features are selected as the most predictive. Regardless of our hyperparameter choices, the top three features are the physics GRE score, achievement orientation, and quality of work. However, the pattern becomes less diagonal toward the bottom right, suggesting that these features are more or less noise in the model.

\subsection{Using Tomek Links to Better Model the Data}

\begin{table}[]
    \centering
    \caption{Metrics when using Tomek Links and MICE for each of the three data sets}
    \begin{tabular}{c|c|c|c} \hline
         & Data Set 0 & Data Set 1a & Data Set 1b \\ \hline
         Cases Dropped & 11\%-14\% & 15\%-18\% & 12\%-17\% \\
         Training AUC & 0.880 $\pm$ 0.004 & 0.760 $\pm$ 0.015 & 0.779 $\pm$ 0.010 \\
         Testing AUC & 0.809 $\pm$ 0.009 & 0.670 $\pm$ 0.015 & 0.704 $\pm$ 0.014 \\
         Testing Accuracy & 0.806 $\pm$ 0.009 & 0.775 $\pm$ 0.012 & 0.717 $\pm$ 0.012 \\
         Null Accuracy & 0.539 $\pm$ 0.006 & 0.699 $\pm$ 0.009 & 0.575 $\pm$ 0.010 \\ \hline
    \end{tabular}
    \label{tab:tomek_table}
\end{table}

Given the limited ability of the conditional inference forest to model data sets 1a and 1b, we used Tomek Links to remove boundary cases. As the goal was to build models that better fit the data, we focused on the model metrics instead of importances. The results are shown in Table \ref{tab:tomek_table}. As MICE generates new values for each imputation and hence, affects which cases are nearest neighbors, the percent of cases dropped for each trial varies.

First, we notice that for data set 0, using Tomek Links increased the testing AUC and testing accuracy by 0.05 over the original model. The testing AUC is now about 0.8 which is considered ``good'' compared to ``fair'' for the original model \cite{araujo_validation_2005}.

Likewise, using Tomek Links also results in an approximately 0.05 increase in the testing AUC and testing accuracy for data set 1a. However, the AUC is still in the poor range and the testing accuracy is only slightly better than the null accuracy.

For data set 1b, using Tomek Links increases the testing AUC and testing accuracy by approximately 0.04. This time, the increase to the testing AUC is enough for the model to be classified as ``fair''.

To better understand what Tomek Links were doing in the modeling process, we investigated how removing the boundary cases affected the decision boundary. To plot the results, we only used the physics GRE score and undergraduate GPA to make a simple model for data sets 0 and 1a. To compute the Tomek Links, we used MICE to create a complete data set first and then found the Tomek Links. As all the data in data set 1b was categorical, a 2D plot of the decision boundary would have yielded limited insight and hence, we did not do so. The results of a single trial are shown in the supplemental material.

For both cases, we find that using Tomek Links appears to reduce the overfitting. Applicants with higher physics GRE scores and higher GPAs were predicted to be admitted while applicants with lower physics GRE scores and GPAs were predicted not to be admitted.

For the feature importances, we find that the ordering of the features is more or less the same as presented in Figures \ref{fig:pre_auc_imp}, \ref{fig:post_auc_imps}, and \ref{fig:rubric_auc_imp}. The plots are included in the supplemental material.

\subsection{Understanding Whether the Admissions Process Changed by Using a True Testing Set}
% \begin{table}[t]
%     \caption{Metrics when applying the pre-rubric model to the post-rubric data and vice versa}
%     \label{tab:prepost_models}
%         \begin{subtable}{.5\linewidth}
%       \centering
%         \caption{Pre-rubric model on post-rubric data}
%         \begin{tabular}{ll} \hline
%         Metric         & Value \\ \hline
%         Pre-Rubric Test AUC & 0.7534 $\pm$ 0.0013 \\
%         Pre-Rubric Test Accuracy & 0.7534 $\pm$ 0.0013 \\
%         Pre-Rubric Null Accuracy & 0.5275 $\pm$ 0.0006 \\ \hdashline
%         Post-Rubric Test AUC & 0.6482 $\pm$ 0.0003 \\
%         Post-Rubric Test Accuracy & 0.6406 $\pm$ 0.0004 \\
%         Post-Rubric Null Accuracy & 0.6653 \\ \hline
%         \end{tabular}
%     \end{subtable}%
%     \begin{subtable}{.5\linewidth}
%       \centering
%         \caption{Post-rubric model on pre-rubric data}
%         \begin{tabular}{ll} \hline
%         Metric         & Value \\ \hline
%         Post-Rubric Test AUC & 0.6301 $\pm$ 0.0010 \\
%         Post-Rubric Test Accuracy & 0.7141 $\pm$ 0.0011  \\
%         Post-Rubric Null Accuracy & 0.6601 $\pm$ 0.0009 \\ \hdashline
%         Pre-Rubric Test AUC & 0.6410 $\pm$ 0.0006 \\
%         Pre-Rubric Test Accuracy & 0.6559 $\pm$ 0.0006  \\
%         Pre-Rubric Null Accuracy & 0.5234 \\ \hline
%         \end{tabular}
%     \end{subtable}
% \end{table}

\begin{figure*}
    \centering
    \includegraphics[width=.95\linewidth]{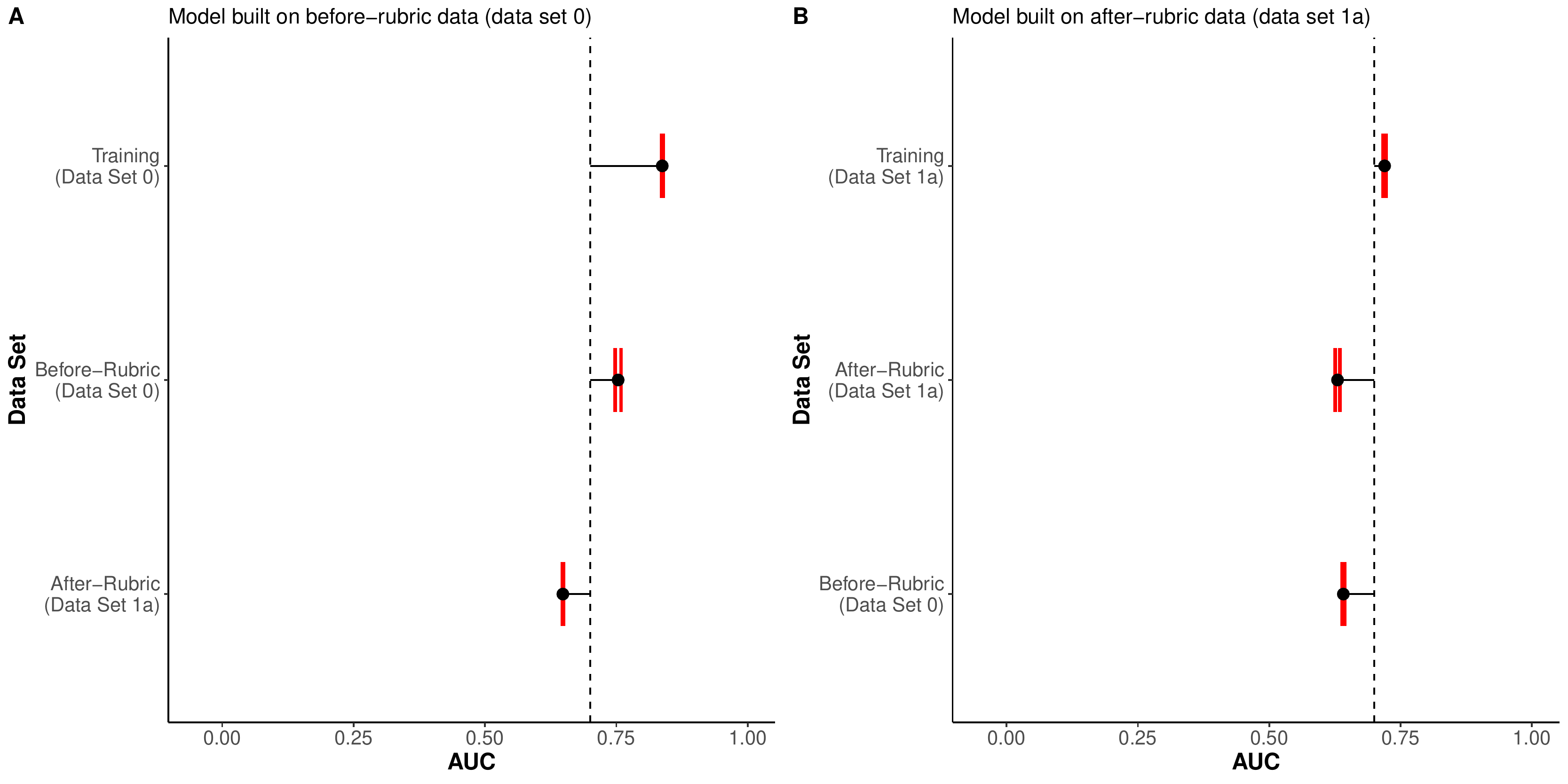}
    \caption{Comparison of the testing AUC when A) Data Set 0 is used to train the model and B) when Data Set 1a is used to train the model. Training refers to the training AUC for the model. All error bars are 1 standard error. Results were averaged over 30 trials.}
    \label{fig:pre_post_auc}
\end{figure*}

\begin{figure*}
    \centering
    \includegraphics[width=.95\linewidth]{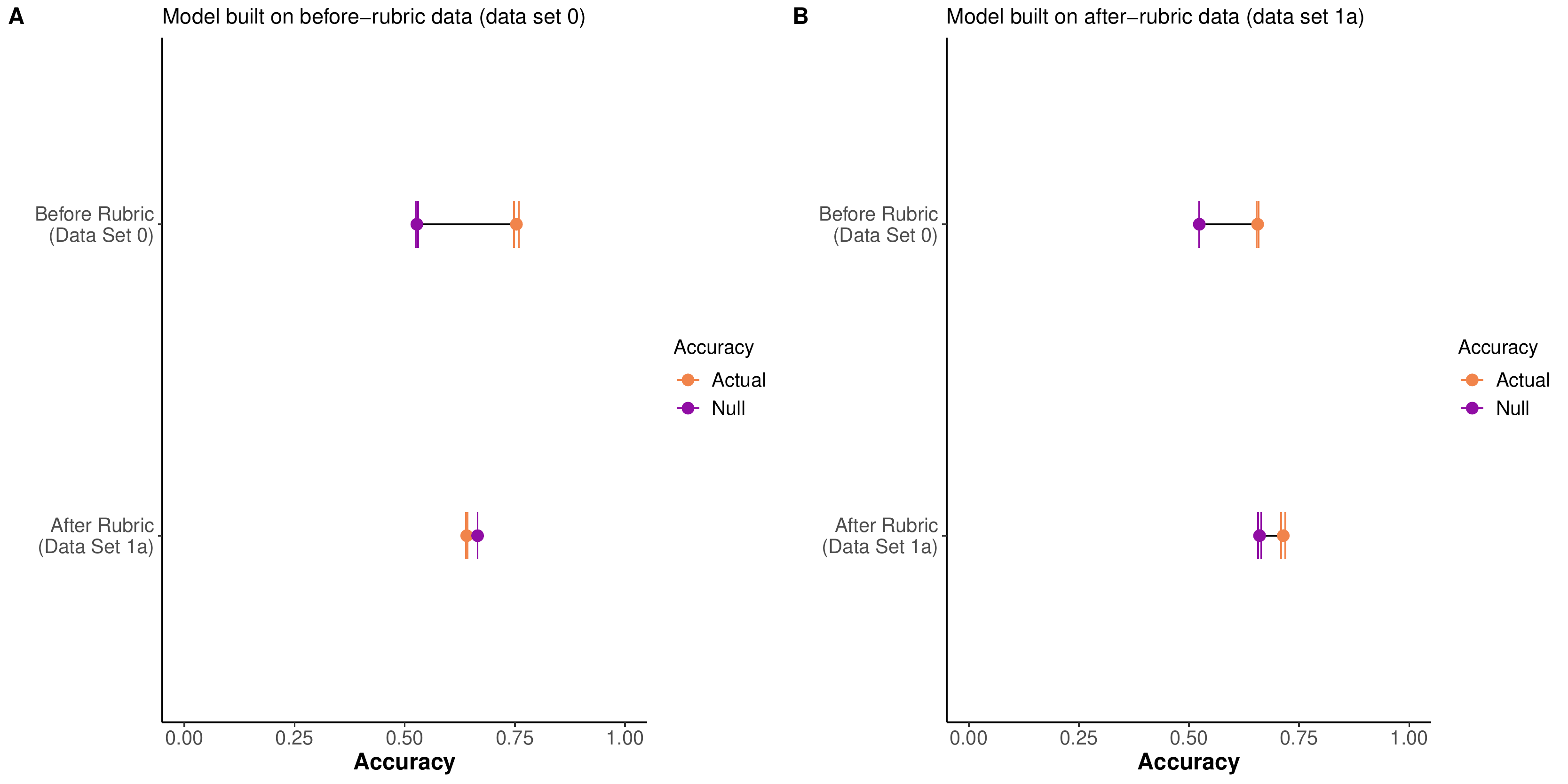}
    \caption{Comparison of the testing accuracy when A) Data Set 0 is used to train the model and B) when Data Set 1a is used to train the model. The null accuracy is shown in cyan with the shorter in height error bars. All error bars are 1 standard error. Results were averaged over 30 trials.}
    \label{fig:pre_post_acc}
\end{figure*}

When looking at the results, which are shown in Figures \ref{fig:pre_post_auc} and \ref{fig:pre_post_acc}, we see that models built on one data set do not work sufficiently well on the other. In Figure \ref{fig:pre_post_auc}A, we see that the data set 0 test AUC is larger than the data set 1a AUC, and in Figure \ref{fig:pre_post_acc}A, we see that the data set 0 test accuracy is larger than the data set 0 null accuracy while the data set 1a test accuracy is smaller than the data set 1a null accuracy. These metrics suggest that the data set 0 model fits data set 0 well but does not fit data set 1a well and therefore, that the process might have changed.

\begin{table}[ht]
\caption{F-statistics and corrected p-values for predicting on each data set and the metric used to assess whether the predictions of the models built on data sets 0 or 1a were different}
\centering
\begin{tabular}{llcr}
  \hline
Data tested on & Metric & F & corrected p-value \\ 
  \hline
Data Set 0 & AUC & 18.95 & 0.01 \\ 
Data Set 0 & Accuracy & 9.70 & 0.03 \\
Data Set 1a & AUC & 1.54 & 0.33 \\ 
Data Set 1a & Accuracy & 4.14 & 0.13 \\ 

   \hline
\end{tabular}
\label{tab:cv52_results}
\end{table}

Looking at Figure \ref{fig:pre_post_auc}B, we see that none of the metrics are especially good. The test AUCs are both in the poor range, suggesting that the model built from data set 1a does not fit that well in the first place. It is then not surprising that the model does not predict data set 0 well. Given that the initial model did not fit the data well, we cannot use the result to make a claim about whether the process changed.

When we look at the 5x2 cv combined F-test, which compared a model for admission before the implementation of the rubric to a model for admission after the implementation of the rubric on the two data sets using two performance metrics, we see similar results (Table \ref{tab:cv52_results}). We find that the models for before and after the implementation of the rubric tested on the before data set (data set 0) are statistically different while the models for before and after the implementation of the rubric tested on the after data set (data set 1a) are not. However, given the results presented in Figures \ref{fig:pre_post_auc} and \ref{fig:pre_post_acc}, the lack of statistical differences for the models tested on data set 1a is likely because both models were equally bad at fitting the data rather than a similar underlying admission process captured by the models.

\section{Discussion}
Here, we first provide answers to our research questions and then use those answers to address the larger question of whether our department's admissions process changed.

\subsection{Research Questions}
\emph{How do admissions models before and after the implementation of the rubric compare in terms of predictive ability and meaningful features when our models are based on the data contained in applications?}

While we were able to model the data before the implementation of the rubric to an acceptable degree, we were unable to do so for the data after the implementation of the rubric. Even after hyperparamter tuning, we were unable to achieve a testing accuracy more than a few percentage points above the null accuracy or a testing AUC above 0.7, suggesting a poor model.

In terms of the meaningful features for data sets 0 and 1a, they were more or less the same. For data set 0 and our process before the implementation of the rubric, we found the applicant's physics GRE score, quantitative GRE score, and GPA to be the meaningful features, while for data set 1a and our admission process after the implementation of the rubric, we found the physics GRE score, GPA, quantitative GRE score, verbal GRE score, and proposed research area to be meaningful. After taking correlations into account, only the physics GRE score and proposed research area were found to be meaningful. The general result of quantitative metrics being most important to the admissions process aligns with previous work that examined the process from the perspectives of faculty \cite{posselt_inside_2016, potvin_investigating_2017}.

Despite prior work suggesting institutional characteristics play an important role in graduate admissions, we did not find institutional or departmental characteristics to be meaningful to models of data set 0 or data set 1a. Our result could be due to differences in methodology or due to institutional effects being influential but not dominant factors \cite{attiyeh_testing_1997}. Indeed, Posselt suggests institutional factors might be used to differentiate applicants with similar GPAs and GRE scores \cite{posselt_inside_2016}. Therefore, we might not have found institutional factors to be meaningful because they are used when primary factors such as GPA and GRE scores do not sufficiently separate applicants.

While the verbal GRE score was not found to be a meaningful feature in data set 0 but was in data set 1a, our program appears to place more emphasis on it than the average program. This may be because our study only looked at domestic students while Potvin et al.'s looked at all applicants \cite{potvin_investigating_2017}. Because international students also take the TOEFL while domestic students do not and admissions directors ranked the TOEFL as more important than the verbal GRE, the TOEFL may take the place of the verbal GRE and hence lower the perceived value of the verbal GRE relative to other factors.

In any case, because conditional inference forests will always return importance values regardless of how well the model fits, we should interpret the data set 1a results with a degree of caution. 

When using a model built on one data set on the other, we found that the model trained on data set 0 did not predict data set 1a well while the model trained on data set 1a did not predict either of the data sets well. We obtained a similar result by our 5x2 cv combined F-test where the predictive accuracy and AUC between models built on data sets 0 and 1a differed on data set 0 and the result was statistically significant. These results suggest that the underlying models are different.

\emph{How does using the data produced by faculty when rating applicants using the rubric affect our ability to create admissions models?}

While using the rubric features does result in improved metrics compared to the traditional features for the data collected after the implementation of the rubric, the metrics are still outside of the acceptable range. The testing AUC was still below 0.7, but the testing accuracy was greater than the null accuracy by a larger amount than the model created from data set 1a. However, that result may be explained by data set 1b having a less imbalanced outcome.

To see if that was the case, we created a model using the data in data set 1a that corresponded to the applicants in data set 1b. When we did so, we found that the metrics were comparable, but the original test data set 1b model slightly outperformed this new model (~0.02 increase in testing AUC and accuracy). Thus, while some of the improvements in metrics might be attributable to the more balanced data set, using the rubric constructs also provided some benefit.

In terms of the features, we noticed some similarities and some differences. For the models of data set 1b, the physics GRE was still the top feature, while measures of the GPA and general GRE scores were ranked in the lower half, suggesting they might not have been as important. Instead, measures of research ability and experience and noncognitive skills tended to be ranked towards the top. Again however, we should interpret the data set 1a results with a degree of caution as the model does not fit the data especially well.

\emph{How does using Tomek Links affect our ability to model the admissions data, both before and after the implementation of the rubric?}

Using Tomek Links resulted in improved model performance for all three data sets. For data set 0, using Tomek Links increased the testing AUC over 0.8, which is considered ``good,'' and for data set 1b, using Tomek Links increased the testing AUC over 0.7, which is considered ``fair.'' However, while using Tomek Links for data set 1a did improve the testing AUC, it did not do so enough for the model to be considered acceptable. 

When looking at the decision boundaries for data sets 0 and 1a with and without Tomek Links removed, we found that overfitting appeared to be reduced, suggesting that even if the metrics are not largely improved, there still may be benefits from using Tomek Links. 

Thus, while the benefits were relatively small, these results suggest that Tomek Links are a promising technique for modeling PER data, especially for data sets where we expect many boundary cases or cases that go against the general trend. For example, if we were to predict who passes an introductory class, Tomek Links might allow us to remove students who earned exam scores around the minimum passing grade and thus might or might not have passed the course or anomalous students who did poorly on the midterms but managed to earn a high grade on the final to pass the class.

\subsection{Addressing whether our process changed}
Looking across the research questions, we can now address our larger question of \textit{did the introduction of the rubric changed our department's admissions process}. Overall, the evidence points in the direction of the process changing.

In terms of evidence for the process changing, we find that the models of data sets 1a and 1b do not fit the data well. As we were able to fit the data set 0 models to an acceptable degree using the conditional inference forest algorithm but not the models of data sets 1a or 1b, this result seems to imply that there must be something different about the data sets. Because data set 0 and data set 1a used the same features, it is hard to explain why we could model one well but not the other unless the ``true'' models of the data were different and hence, the admission process changed.

In addition, a model trained on data set 0 was better able to predict held-out data from data set 0 compared to data set 1a and the 5x2 cv combined F-test found statistically significant differences in the performance of the models. If the process had not changed, we would have expected the predictive performance to be similar.

Finally, using Tomek Links to remove applicants who might have gone against the general trend resulted in minimal increases in the metrics for the models of data sets 1a and 1b. If the process did not change, we would expect that removing applicants who might have gone against the overall trend would have led to a better model because we were able to model the admissions data before the implementation of the rubric. Yet, that isn't what happened, suggesting again there must be something different about the data collected after the implementation of the rubric.

\subsection{Limitations affecting our ability to address whether the process changed}
Looking at the results, it is possible that someone could instead believe the results suggest the process did not change. We address those here.

In terms of evidence for the process not changing, our results show that the most predictive features are similar regardless of which data set we used. When using data set 0, we found that the physics GRE, quantitative GRE, and GPA were most predictive of admission. Likewise, when looking at data set 1a, we found that the physics GRE, GPA, quantitative GRE, verbal GRE, and proposed area of research were the most predictive. Using data set 1b showed the most differences in that the measures of grades and the general GRE scores were in the lower half of the rankings. However, the physics GRE was still the top ranked feature. Yet, both models of the data after the implementation of the rubric did not have acceptable testing metrics, suggesting that we should interpret the feature importance orders with caution. Conditional inference forest models will always produce feature importances regardless of how well the model fits the data. Because the metrics to assess fit are relatively poor, we should not trust the conclusion that the most predictive features are the same between these models.

However, it is possible that the low metrics might be a result of the conditional inference forest method not being suited for the data we have. Recent work suggests that the conditional inference forest algorithm does not perform well with missing data \cite{hu_variable_2021}. When we used MICE to impute the missing data, the models were still not able to produce testing metrics in the acceptable range, suggesting that the missing data was not the issue. In addition, a recent study using admissions data to predict later performance in a graduate program found that random forest methods were among the best performing methods compared to other common methods such as logistic regression, support vector machines, Naive Bayes, and neural networks, suggesting that our choice of algorithm is unlikely to be creating the observed poor performance \cite{zhao_predicting_2020}.

In addition, while conditional inference forests were designed to better handle categorical data than traditional random forests do, there could still be issues with categorical data. For example, for data set 1b, there are only three possible values for each feature. Therefore, the model can only split each feature 3 ways, which limits the depth of the trees and the fine tuning of the model. However, when we used the section total (which could take on any integer between 0 and 8), the results did not substantially improve, suggesting that the scale of the data may not be to blame.

Even if the number of categories does not matter, the fact that some of the categorical data are discretized, continuous features (e.g., physics GRE score, physics coursework) could create problems. Prior work has shown that binning continuous features can lead to a loss of information and over- or under-estimation of effect sizes \cite{maccallum_practice_2002,irwin_negative_2003}. It is possible that such an effect is present in our data. However, models built from data sets 1a and 1b both found the physics GRE score to be the top feature even though the physics GRE score was discretized in data set 1b. Because the model metrics were not great (the testing accuracy was only a few percentage points above the null accuracy an the testing AUC was less than 0.7), this rebuttal should be treated with caution. On the other hand, the fact that models of data set 1a, where discretization was not an issue, still had poor metrics suggests that it cannot fully explain the models' low metrics.

It is also possible that the low metrics are not a result of how we handled the data we had, but rather what data we had. It is possible that the applicant pools differed substantially before and after the implementation of the rubric or that committee members were using something not included in our data to evaluate applicants and if we had that data, our models of data set 1a and 1b would improve. An analysis of the applicant pools (included in the appendix), suggests the applicant pools are not substantially different on key measures and while such an explanation about extraneous features seems possible for data set 1a, it seems unlikely for data set 1b because members of the department decided what qualities they wanted to evaluate applicants on and added them to the rubric.

In addition, it is possible that data sets we had were too small for us to properly model. That is, if data sets 1a and 1b were larger, perhaps we would have been able to produce models with acceptable testing metrics and hence, trust the importance rank results. However, given that data set 0 and data set 1a were of similar sizes, it would be difficult to explain why we were able to create acceptable models for data set 0 but not data set 1a if the underlying admissions processes were the same.

Finally, it is possible that the low metrics might not be caused by the data or the model and instead, the low metrics could be caused by the admission process itself. The goal of the rubric is to rate applicants along multiple dimensions, and hence in a holistic manner. If applicants were actually assessed holistically, we would expect that the model would not generalize well because there is no single underlying process. Instead there might be multiple routes an applicant could take to gain admission and hence, the model might encounter difficulties modeling this process. The fact that hyperparameter tuning and Tomek Links did not increase the testing metrics to an acceptable range for models of data set 1a and barely did so for the models of data set 1b supports such an interpretation. However, claiming the process is more holistic based on these results alone is premature, especially given the relatively small number of applicants in data set 1b. Instead, results from other modeling attempts would either need to show poor predictive ability or show evidence of multiple routes to admission to support such a claim.

\section{Future Work}
To better address the limitations and consider whether our admissions process became most holistic, future work should examine alternative techniques for analyzing the data.

First, instead of taking a predictive approach in our analysis, we could take an explanatory approach where we try to understand what inputs may have caused the outcome. Under this approach, whether a feature is related to the outcome is determined by statistical significance rather than its predictive ability \cite{shmueli_explain_2010}. Logistic regression is a common example of this technique in PER. The results of such future work would provide greater insight into why the models did not fit data sets 1a and 1b well.

Second, to determine if the process is more holistic, future work could analyze the data using cluster analysis or latent class analysis. While such methods are becoming popular for analyzing learning environments (e.g., see \cite{stains_anatomy_2018,commeford_characterizing_2021}), to our knowledge, such methods are less common in studies of graduate admissions processes. To our knowledge, clustering-like techniques have only been used to understand admissions strategies based on surveys of faculty on admissions committees \cite{doyle_search_2015}. If the process is more holistic, such methods might be able to identify clusters of applicants who were admitted for similar reasons. For example, some applicants may be admitted due to stellar academic credentials, others may be admitted due to their research background, while others may be admitted based on which faculty members are seeking new students. Finding or not finding such a result would provide greater clarity as to how the process may have changed. To do so however, would likely require a larger data set, especially if there are a large number of driving results for why an applicant is admitted.

Finally, future work should take a mixed methods approach by considering qualitative approaches to investigating how our admissions process might have changed. Such qualitative approaches could allow us to observe the admissions process itself (similar to the studies Posselt conducted as documented in \cite{posselt_inside_2016}) and understand how faculty are evaluating and discussing applicants in real time. In addition, a qualitative approach would allow us to avoid many of the modeling limitations related to the scale of the data and metrics.

Alternatively, future work could directly ask faculty who have served on the admissions committee both before and after the implementation of the rubric about their perception of the process at each time. However, we must be careful of faculty's potential biases when recalling how things were done in the past (see Muggenburg for an overview \cite{muggenburg_beyond_2021}). For example, given the greater emphasis on diversity and equity in higher education now, faculty's recall may suffer from post-rationalization \cite{behrens_analysing_2008} where they justify their decisions using reasons that weren't available at the time but are consistent with their current self image or social desirability \cite{edwards_social_1957} where past events may be distorted to conform to current attitudes and norms.

More broadly, future work should consider the admissions process at other physics departments and understand how changes designed to make the process more equitable work in practice at other institutions. This study was done at a primarily white institution (PWI) and might not be applicable to universities with differing applicant populations. While Kanim and Cid note that having a relatively homogeneous research sample can be valuable for reducing variability, especially in early studies, they also note that exploring the effects of variability can lead to new results and a greater understanding of the results \cite{kanim_demographics_2020}. Thus, while our results might generalize to many physics graduate programs, it might also hide important differences in features predictive of admission for applicants of different demographics groups and institutions with different demographics than our own.

\section{Conclusion}
Overall, the results of this initial investigation are suggestive that our admission process did change after the implementation of the rubric. We were able to model the data from before the implementation of the rubric to a sufficient degree but not the data after the implementation of the rubric. In addition, the model of the admissions process before the implementation of the rubric does not do well predicting the data collected after the implementation of the rubric and vice versa, suggesting that the underlying process did change. However, there are still numerous limitations that need to be addressed before we can make a definitive conclusion, including how we characterize the data and how we model the data.

Furthermore, the models of the data following the implementation of the rubric performing poorly suggests that the process might be holistic. In order to make such a conclusion, however, we would need either evidence in favor of the occurrence holistic admissions or stronger evidence that the current admission process is not easily modeled by known techniques. Such evidence could be obtained through a variety of quantitative or qualitative approaches.

In terms of the modeling approaches, Tomek Links seem like a promising technique for future PER studies. While their use was not enough to provide a more conclusive answer to the question whether our admission process changed, their use did provide evidence that the data collected after the implementation of the rubric may be modelable to an acceptable level, leaving open the possibility that other methods may be able to model the data and hence, should be explored.

Finally, to truly get a sense of whether admission processes change after the implementation of a rubric or merely use a new tool to do the same process, studies such as these need to be completed in other physics departments. By doing so, we will have a better idea of how rubric-based admissions might change admission processes and how well our results generalize to other programs.

\acknowledgments{We would like to thank Scott Pratt, Remco Zegers, and Kirsten Tollefson for providing the data for this project. We would also like to thank Tabitha Hudson for compiling the data. This project was supported by the Michigan State University College of Natural Sciences and the Lappan-Phillips Foundation.}

\section{Appendix: Comparison of data sets}\label{sec:appendix}

An alternative explanation as to why we were able to model the data before the implementation of the rubric (data set 0) but not the data after the implementation of the rubric (data set 1a) could be the underlying data, rather than the admissions process, is different. Here, we provide evidence to suggest that that is not the case.

Given the results of Sec \ref{sec:pre_results} where we could model the data, we compared the distributions of the top features from those models (data set 0) to the distributions of those features from the data after the implementation of the rubric (data set 1a). If the distributions of the features were statistically the same for the two data sets, it is would be difficult to explain why we could model those distributions for data set 0, but not data set 1a.

\begin{figure}[!b]
    \centering
    \includegraphics[width=.95\linewidth]{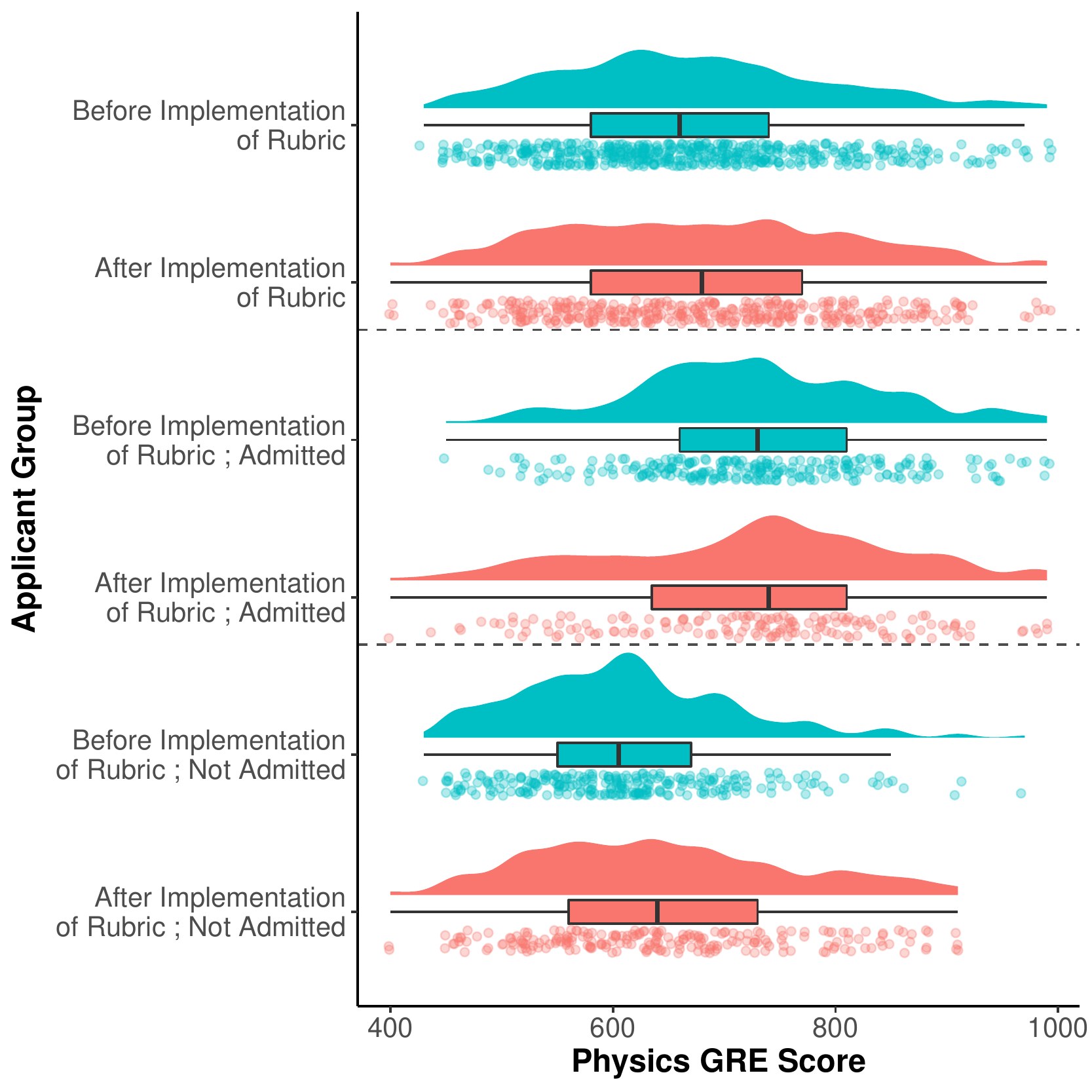}
    \caption{Raincloud plots showing the distribution of physics GRE scores of all applicants before and after the implementation of the rubric, only admitted applicants, and only non-admitted applicants. Only the distributions of physics GRE scores for non-admitted applicants before and after the implementation of the rubric were found to be statistically different.}
    \label{fig:raincloud_pgre}
\end{figure}

\begin{figure}
    \centering
    \includegraphics[width=.95\linewidth]{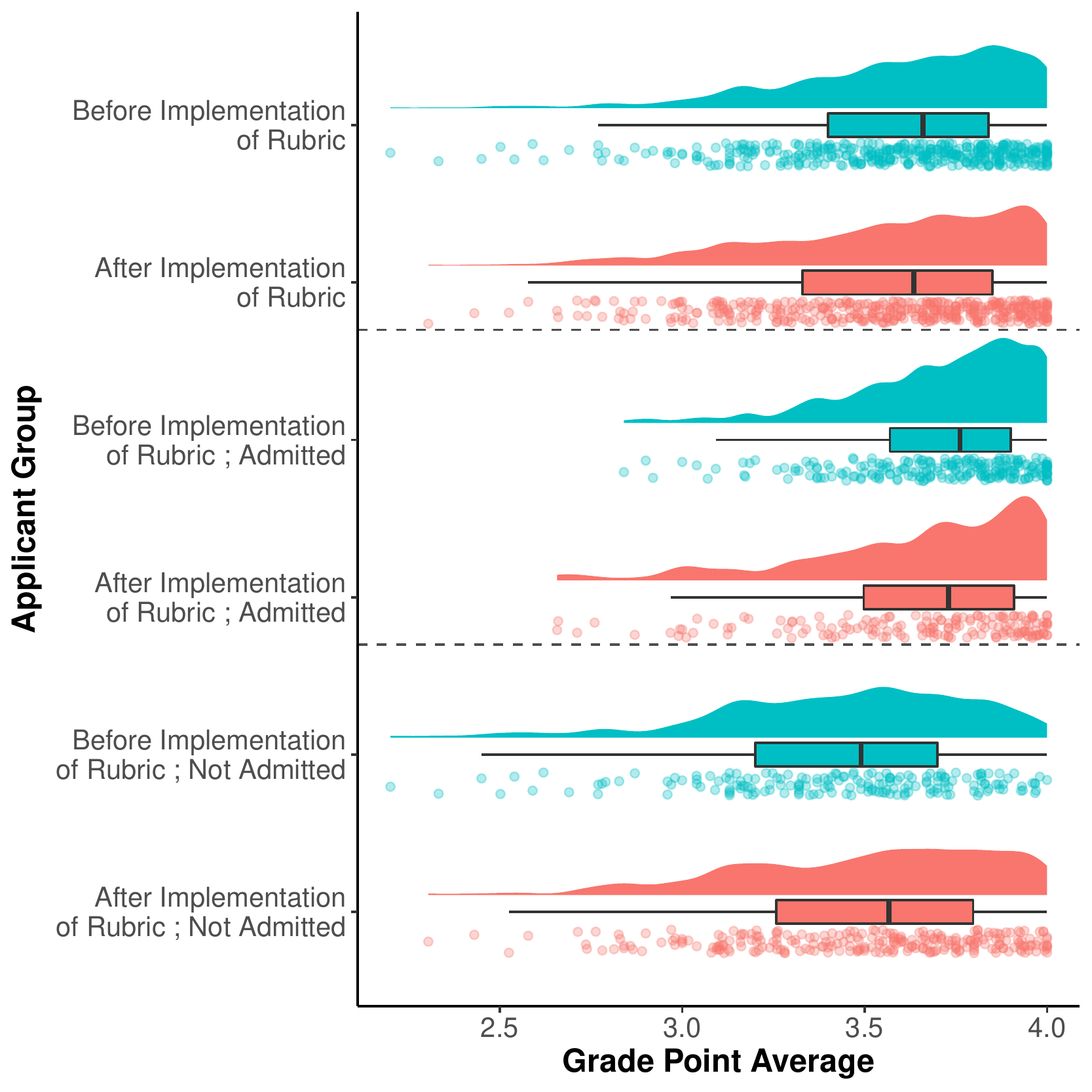}
    \caption{Raincloud plots showing the distribution of grade point averages of all applicants before and after the implementation of the rubric, only admitted applicants, and only non-admitted applicants. None of the distributions of GPAs for applicants before and after the implementation of the rubric were found to be statistically different.}
    \label{fig:raincloud_gpa}
\end{figure}

\begin{figure}[!ht]
    \centering
    \includegraphics[width=.95\linewidth]{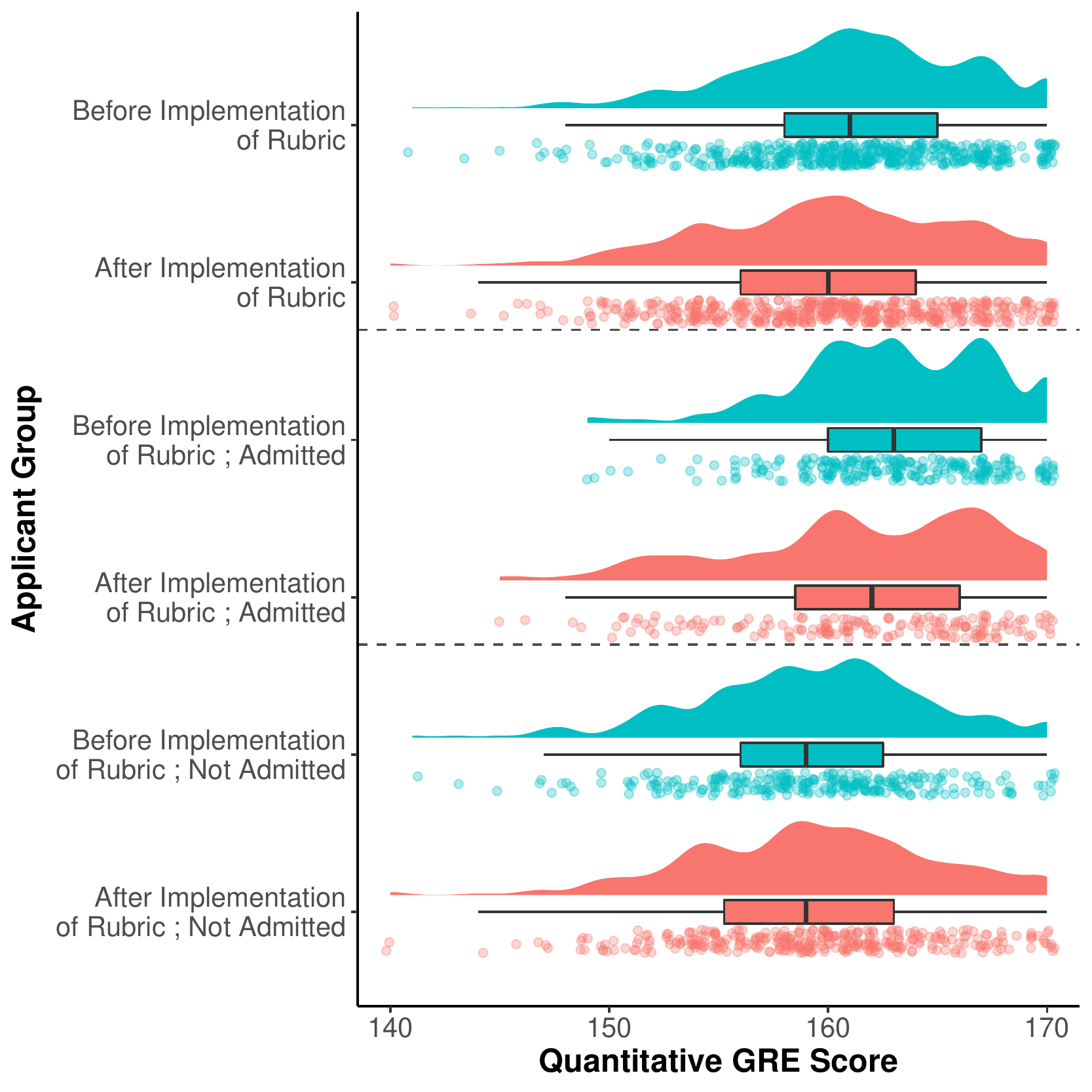}
    \caption{Raincloud plots showing the distribution of quantitative GRE scores of all applicants before and after the implementation of the rubric, only admitted applicants, and only non-admitted applicants. None of the distributions of quantitative GRE scores for applicants before and after the implementation of the rubric were found to be statistically different.}
    \label{fig:raincloud_qgre}
\end{figure}

The raincloud plots \cite{allen_raincloud_2019} of the distributions of the applicant's physics GRE scores, GPA, and quantitative GRE scores, the most predictive features of data set 0, are shown in Figures \ref{fig:raincloud_pgre}, \ref{fig:raincloud_gpa}, and \ref{fig:raincloud_qgre}.

\begin{table*}[]
\caption{D and uncorrected p-value from Kolmogorov-Smirnov test on distributions of applicants before and after the implementation of the rubric}
\label{tab:ks_table}
\begin{tabular}{llcrr}
Feature                           & Group        & D     & uncorrected p-value & significant?\\ \hline
\multirow{3}{*}{Physics GRE}      & All          & 0.080 & 0.130   & no\\
                                  & Admitted     & 0.101 & 0.286   & no\\
                                  & Non-admitted & 0.173 & 0.002   & yes\\
\multirow{3}{*}{GPA}              & All          & 0.064 & 0.371   & no\\
                                  & Admitted     & 0.091 & 0.404   & no\\
                                  & Non-admitted & 0.118 & 0.109   & no\\
\multirow{3}{*}{Quantitative GRE} & All          & 0.091 & 0.032   & no\\
                                  & Admitted     & 0.128 & 0.077   & no\\
                                  & Non-admitted & 0.037 & 0.989   & no\\\hline \hline
\end{tabular}
\end{table*}

From Figure \ref{fig:raincloud_pgre}, we notice that the distributions of the physics GRE scores of all applicants before and after the implementation of the rubric seem similar although the applicants after the implementation of the rubric seem to have a slightly higher median physics GRE score. The admitted applicants after the implementation of the rubric also seem to have a similar median physics GRE score as the admitted applicants before the implementation of the rubric while the non-admitted applicants after the implementation of the rubric had a higher median physics GRE score than the non-admitted applicants before the implementation of the rubric.

In Figure \ref{fig:raincloud_gpa}, we see a similar result when comparing the grade point averages of applicants before and after the implementation of the rubric as well as when we break applicants into admits and non-admits.

In figure \ref{fig:raincloud_qgre}, we see that applicants after the implementation of the rubric had a lower median quantitative GRE score than applicants before the implementation of the rubric. The same is true for admitted applicants while non-admitted applicants had similar median quantitative GRE scores, regardless of whether they applied before or applicants after the implementation of the rubric.

To determine if these differences were statistically significant, we conducted Kolmogorov-Smirnov tests between the applicants who applied before and after the implementation of the rubric \cite{massey_jr_kolmogorov-smirnov_1951}. As there were nine tests (physics GRE score, GPA, and quantitative GRE score for all, admitted, and non-admitted applicants), we used the Holm-Bonferroni method to correct p-values for multiple comparisons \cite{holm_simple_1979}. With this method, the smallest p-value is compared to $0.05/n$, the next smallest p-value to $0.05/(n-1)$ and so on until the null hypothesis is not rejected. At that point, we are unable to reject any remaining null hypotheses.

The results of the Kolmogorov-Smirnov tests are shown in Table \ref{tab:ks_table}. We find that the distributions of physics GRE scores are statistically different for non-admitted applicants before the implementation of the rubric and non-admitted applicants after the implementation of the rubric. For all other comparisons, we are unable to reject the null hypothesis that the distributions are the same. 

Given that two of the three top features for predicting which applicants would be admitted before the implementation of the rubric were not found to have different distributions for any of the groups and the third was only found to have a differing distribution for one of the three groups, it seems that the data is not the reason for our inability to model data set 1a.

To further check is this claim, we reran the model on data set 0 without using applicant's physics GRE score but all of the other features listed in Table \ref{tab:pre_factors}. When we did so, we found a testing accuracy of $0.722 \pm 0.005$ and a testing AUC of $0.722 \pm 0.005$, suggesting a decent model still. Therefore, even though the distribution of physics GRE scores for non-admitted applicants before and after the implementation of the rubric are different, that we are still able to model the data set 0 well enough without the physics GRE scores included suggests that the differences in distributions should not affect our ability to produce a decent enough model of data set 1a.

\bibliography{references.bib}
\end{document}

\title{Supplemental Material to Rubric-based holistic review represents a change from traditional graduate admissions approaches in physics}
\maketitle

Here we present additional figures related to hyperparameter tuning and Tomek links. 

\section{Hyperparameter tuning}
These figures show the results of 125 hyperparameter combinations and the fraction of the combinations in which each figure had the specified rank. A higher rank is meant to mean the feature is more predictive of being admitted to our program.

\begin{figure*}
\centering
  \includegraphics[width=.95\linewidth]{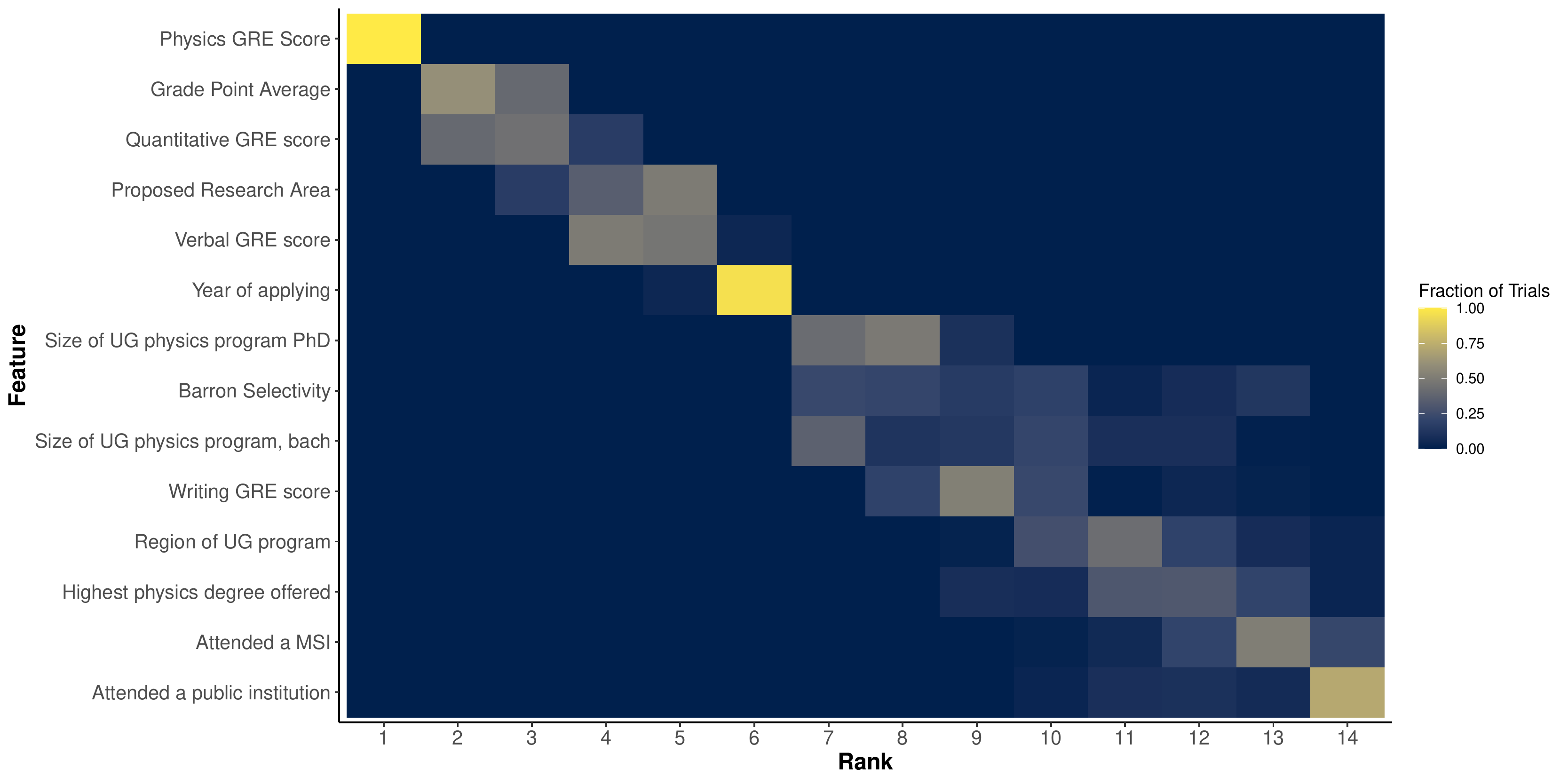}
  \caption{Proportion of the 125 hyperparameter combinations in which each feature had a given rank for data set 0. Notice that there is a block of features that range between 1 and 5 and a block of features that rank between 7 and 14. \label{fig:pre_hyperparameter_rank}}
\end{figure*}

\begin{figure*}
\centering
  \includegraphics[width=.9\linewidth]{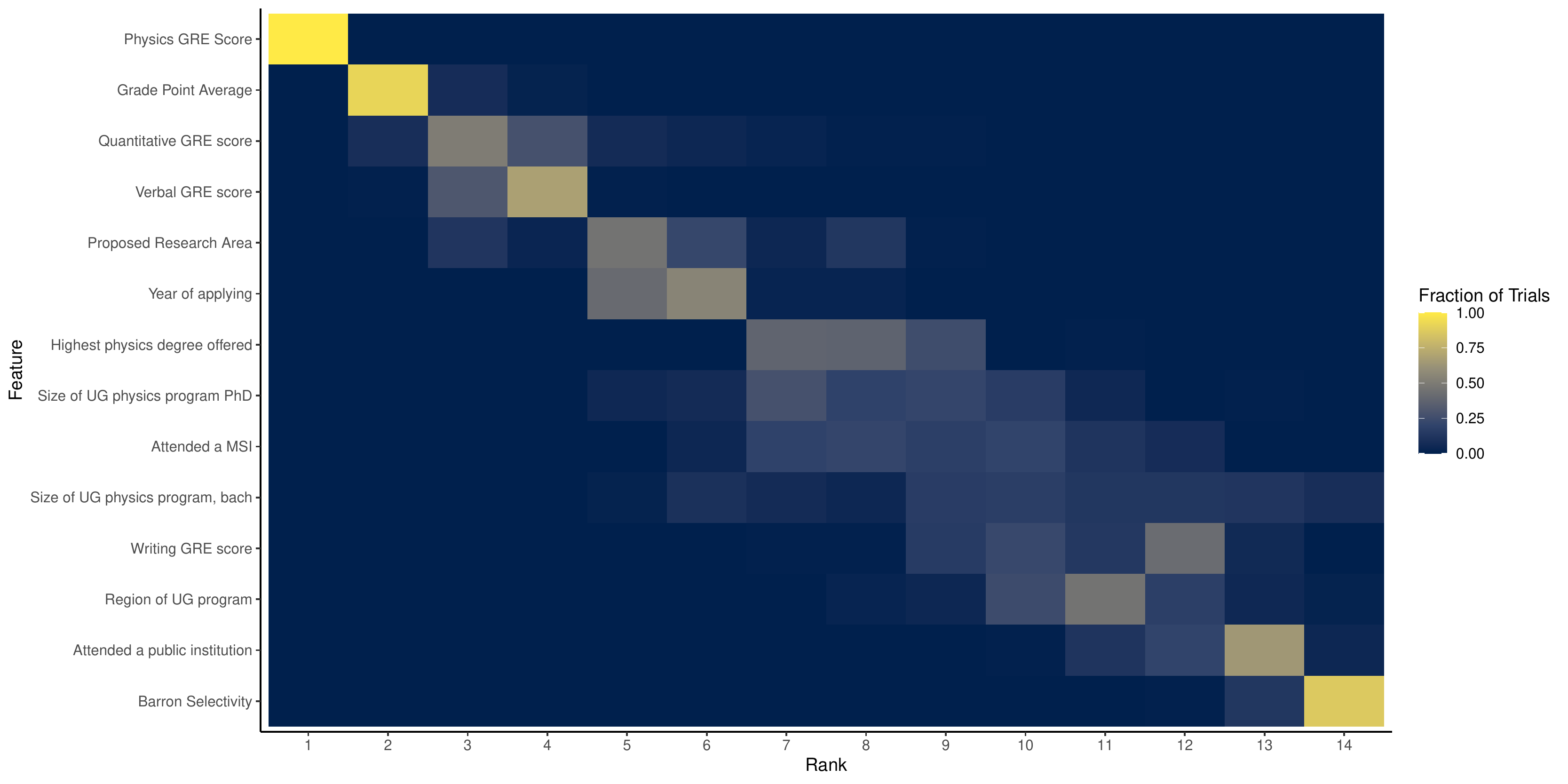}
  \caption{Proportion of the 125 hyperparameter combinations in which each feature had a given rank for data set 1a. Notice that the plot is mostly diagonal and that physics GRE score and GPA are almost always the top two features. \label{fig:post_hyperparameter_rank}}
\end{figure*}

\begin{figure*}
\centering
  \includegraphics[width=.95\linewidth]{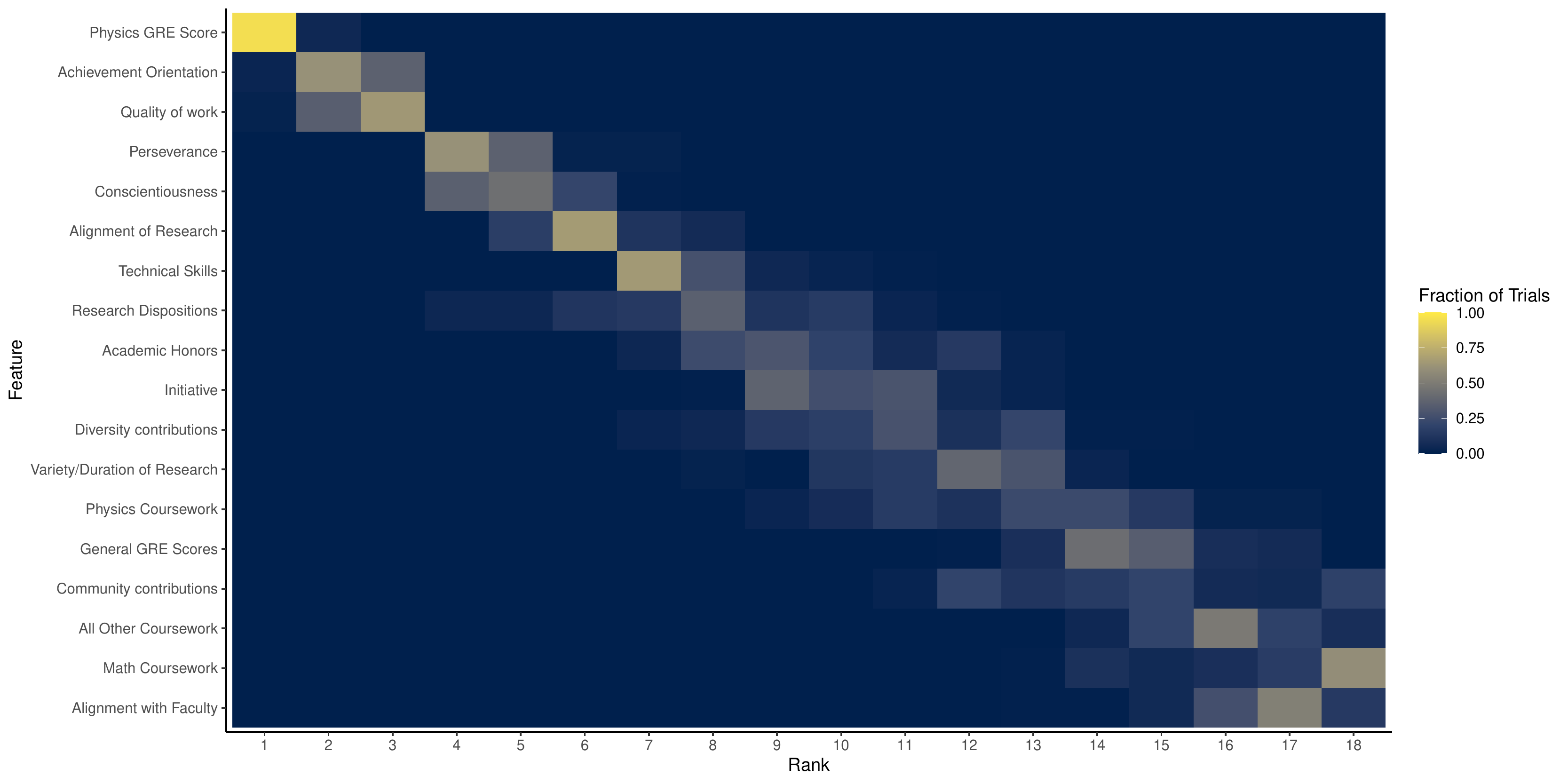}
  \caption{Proportion of the 125 hyperparameter combinations in which each feature had a given rank for models of data set 1b. Notice that the plot is mostly diagonal and that physics GRE score, achievement orientation, and quality of work are always the top three features. \label{fig:rubric_hyperparameter_rank}}
\end{figure*}

\section{Tomek Links}
\begin{figure*}
    \centering
    \includegraphics[width=.95\linewidth]{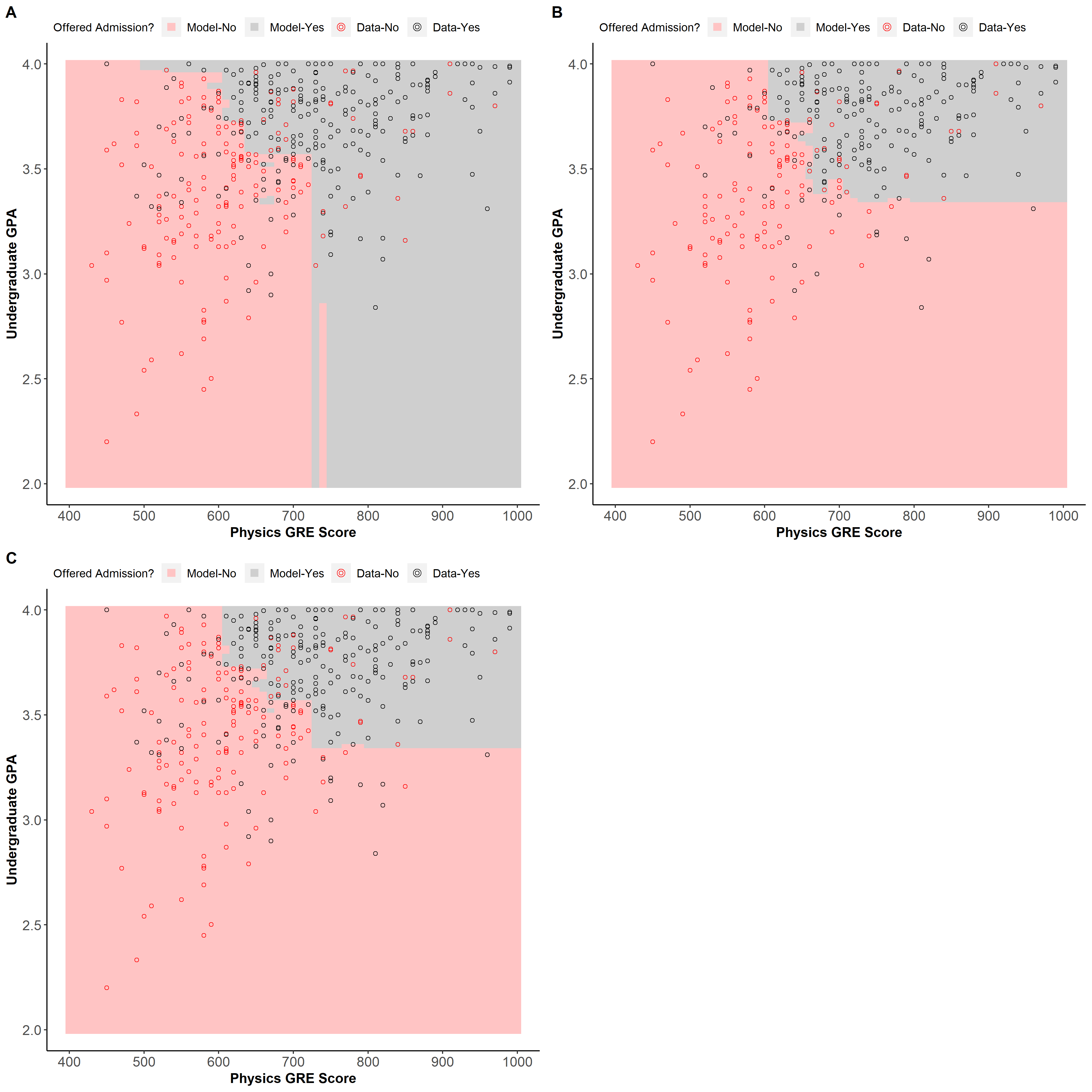}
    \caption{Plot A shows data set 0 with the decision boundary for a model with just the physics GRE score and undergraduate GPA while plot B shows the data with the Tomek Links removed and the resulting decision boundary for the 2D model. Plot C shows the overlap of the admitted regions from plots A \& B.}
    \label{fig:dataset0_boundary}
\end{figure*}

From the figures, we see that removing the Tomek Links does affect the boundary. In the case of Figure \ref{fig:dataset0_boundary}, we see the area with limited data in the lower right switches to not admitted and in general, the overfitting is reduced. In addition, the decision boundary matches closer to what we might expect anecdotally and based on our previous work in that having a higher physics GRE score and GPA is more likely to result in admission as opposed to having only one of those being stellar \cite{young_physics_2021}. In addition, the plot ``A"s suggest an informal cutoff score where applicants scoring less than 700 on the physics GRE are unlikely to be admitted.

\begin{figure*}
    \centering
    \includegraphics[width=.95\linewidth]{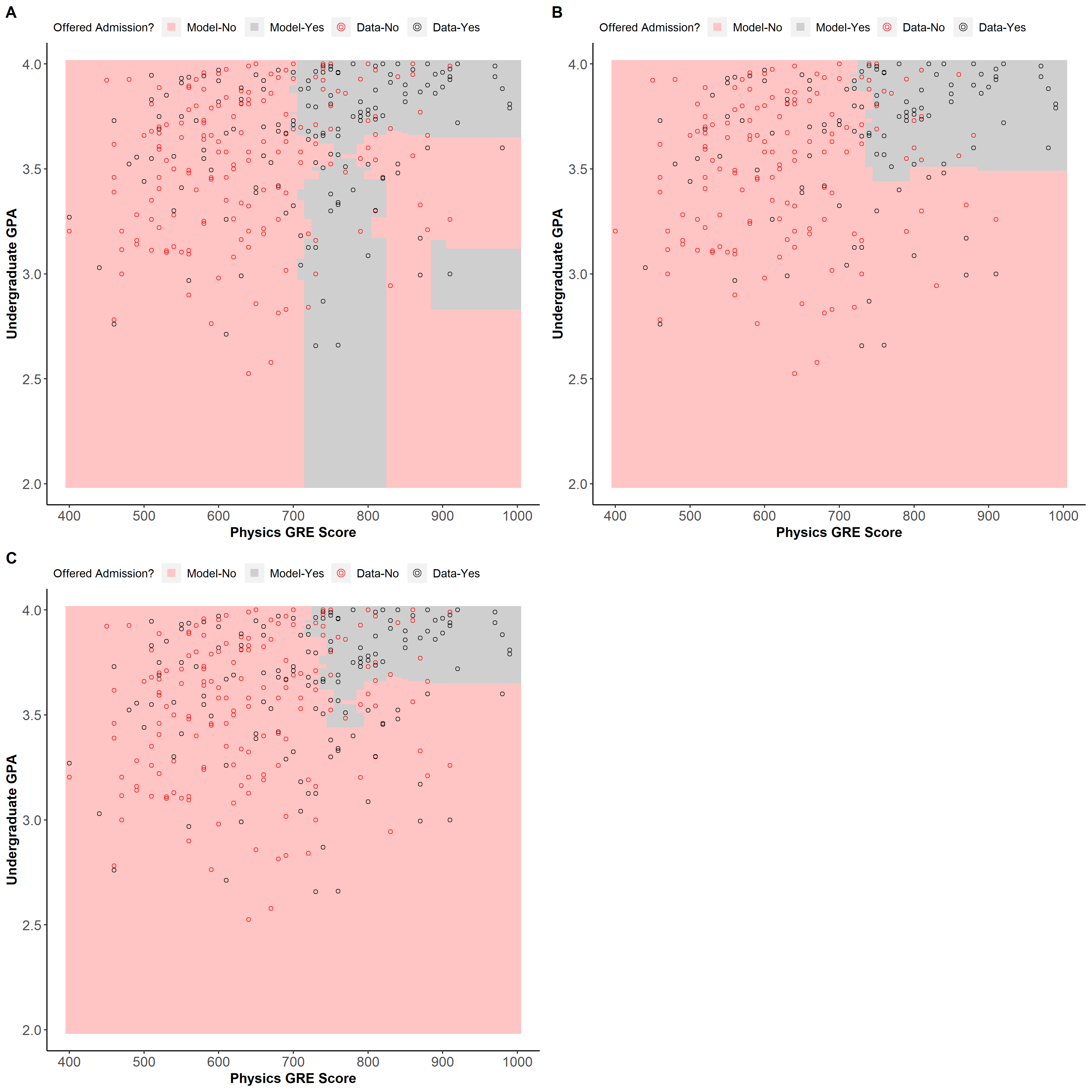}
    \caption{Plot A shows data set 1a with the decision boundary for a model with just the physics GRE score and undergraduate GPA. Plot B shows the data with the Tomek Links removed and the resulting decision boundary for the 2D model. Plot C shows the overlap of the admitted regions from plots A \& B.}
    \label{fig:dataset1a_boundary}
\end{figure*}

Likewise, in Figure \ref{fig:dataset1a_boundary}, we again see reduced overfitting in the decision boundary. We also see that higher physics GRE scores and GPA are predicted to result in admission as was the case before the implementation of the rubric. However, the threshold for what counts as a high physics GRE score and GPA seems to be higher after the implementation of the rubric based on the decision boundaries.

\section{Tomek Links Importance}
Here we include the averaged feature importances. The results are similar to the original data.

\begin{figure}
    \centering
    \includegraphics[width=.5\linewidth]{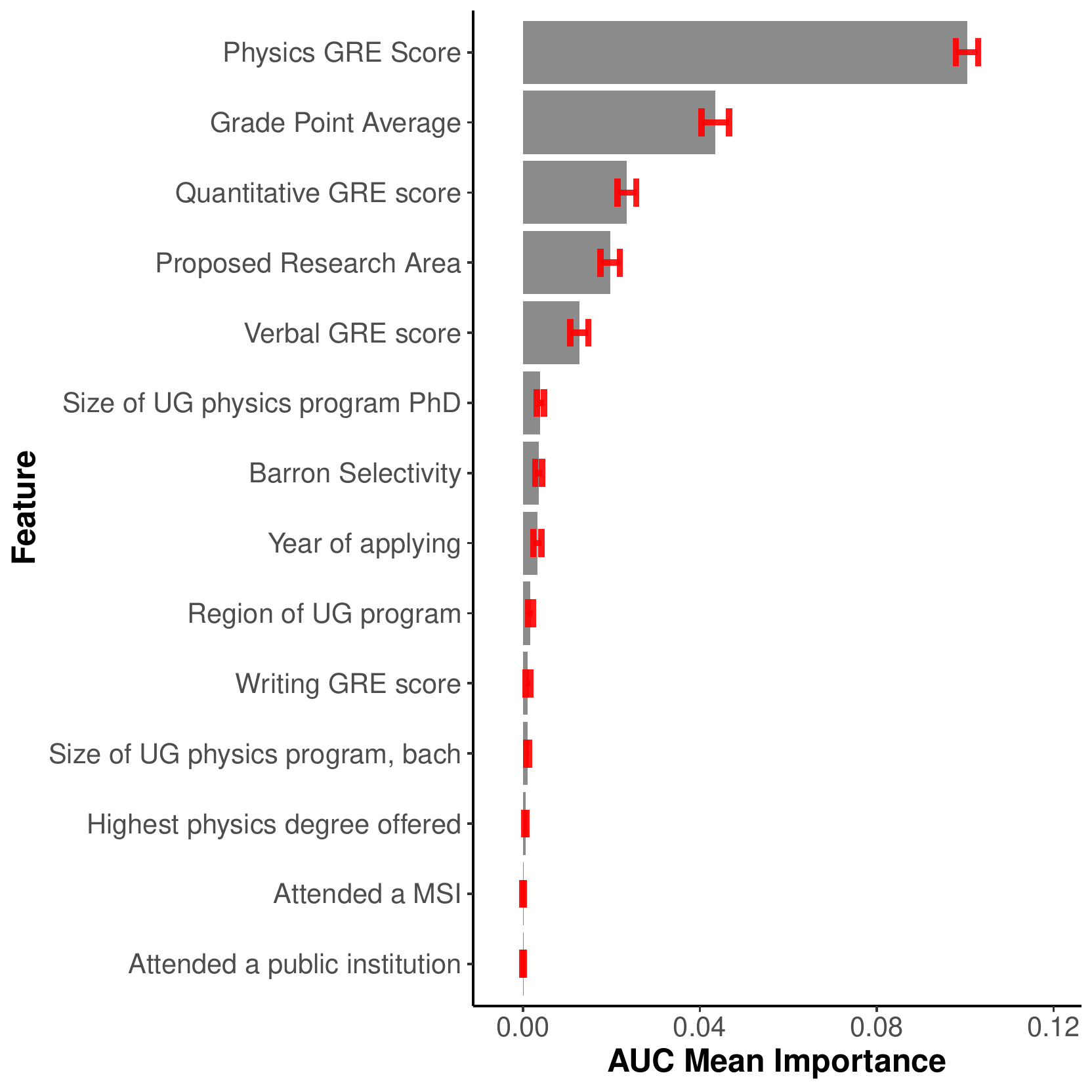}
    \caption{Averaged AUC feature importances over 30 trials for the data collected before the implementation of the rubric with Tomek Links removed (data set 0).}
\end{figure}

\begin{figure}
    \centering
    \includegraphics[width=.5\linewidth]{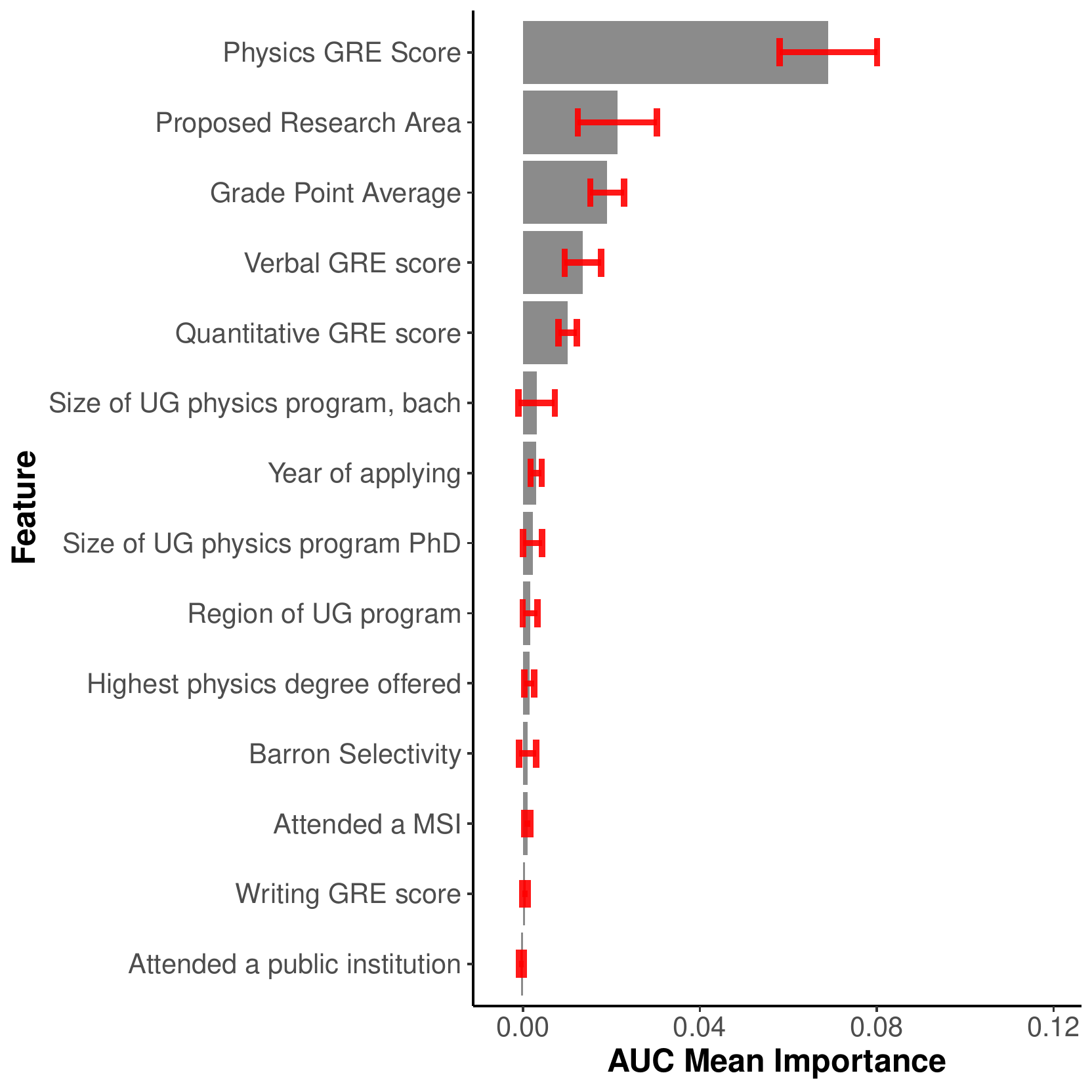}
    \caption{Averaged AUC feature importances over 30 trials for the data collected after the implementation of the rubric Tomek Links removed (data set 1a).}
\end{figure}

\begin{figure}
    \centering
    \includegraphics[width=.5\linewidth]{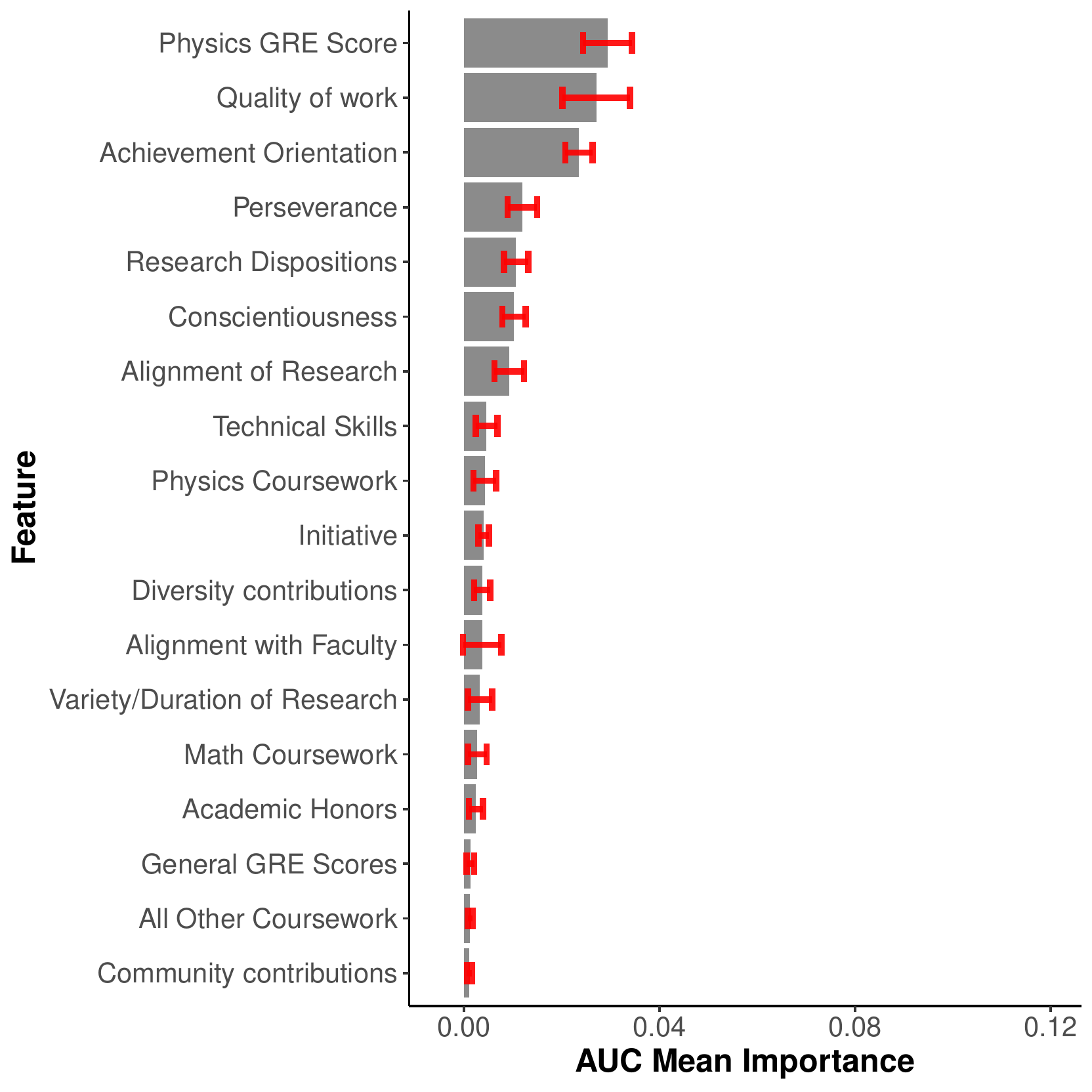}
    \caption{Averaged AUC feature importances over 30 trials for the data collected after the implementation of the rubric Tomek Links removed (data set 1b).}
\end{figure}

\bibliography{references.bib}